\documentclass[12pt]{article}
\usepackage[margin=25mm]{geometry}
\usepackage{amsmath,amsfonts,amssymb}
\usepackage[utf8]{inputenc}
\usepackage[authoryear]{natbib}
\usepackage{graphicx}
\usepackage{verbatim,booktabs,threeparttable,adjustbox,longtable}
\usepackage{pgfplots}
\usepackage{bm}
\usepackage{algorithm}
\usepackage{algpseudocode}
\usepackage{subcaption,caption}
\usepackage{setspace}
\usepackage{tikz}
\usepackage[percent]{overpic}
\usepackage[colorlinks=true, linkcolor=black, citecolor=black, urlcolor=blue]{hyperref}
\pgfplotsset{compat=1.18}
\DeclareMathOperator*{\argmax}{argmax}
\DeclareMathOperator*{\argmin}{argmin}

\providecommand{\keywords}[1]{\small\textbf{Keywords:} #1}

\title{Scalable and Efficient Multiple Imputation for Case-Cohort Studies via Influence Function-Based Supersampling}
\author{
  Jooho Kim\textsuperscript{1} and Yei Eun Shin\textsuperscript{1,2} \\
  \small \textsuperscript{1}Department of Statistics, Seoul National University, South Korea\\[-0.5em]
  \small \textsuperscript{2}College of Liberal Studies, Seoul National University, South Korea
}
\date{}

\begin{document}
\maketitle

\begin{abstract}
Two-phase sampling designs have been widely adopted in epidemiological studies to reduce costs when measuring certain biomarkers is prohibitively expensive. Under these designs, investigators commonly relate survival outcomes to risk factors using the Cox proportional hazards model. To fully utilize covariates collected in phase 1, multiple imputation (MI) methods have been developed to impute missing covariates for individuals not included in the phase 2 sample. However, MI becomes computationally intensive in large-scale cohorts, particularly when rejection sampling is employed to mitigate bias arising from nonlinear or interaction terms in the analysis model. To address this issue, Borgan et al. (2023) proposed a random supersampling (RSS) approach that randomly selects a subset of cohort members for imputation, albeit at the cost of reduced efficiency. In this study, we propose an influence function-based supersampling (ISS) method with weight calibration. The method achieves efficiency comparable to imputing the entire cohort, even with a small supersample, while substantially reducing computational burden. We further demonstrate that the proposed method is especially advantageous when estimating hazard ratios for high-dimensional expensive biomarkers. Extensive simulation studies are conducted, and a real data application is provided using the National Institutes of Health--American Association of Retired Persons (NIH--AARP) Diet and Health Study.
\end{abstract} \hspace{10pt}


\section{Introduction}
In epidemiological studies, two-phase designs have been widely adopted to reduce study costs in situations where measuring certain biomarkers is prohibitively expensive (\citealp{breslow1988logistic, white1982two}). Among these, case-cohort and nested case-control designs are particularly prominent, with well-established methodologies for their analysis (\citealp{prentice1986case, self1988asymptotic, lin1993cox, borgan2000exposure, samuelsen1997psudolikelihood}). In a case-cohort study, a random subcohort is selected, and all cases that experience the event are additionally included, forming the phase 2 sample. An extension of the case-cohort design is a stratified case-cohort design, which samples the subcohort separately within strata. The stratified case-cohort design improves efficiency and ensures representation of rare exposure groups (\citealp{borgan2000exposure}). A nested case-control design proposed by \citet{thomas1977appendix} randomly selects controls from the risk set for each case at its event time. In this study, we adopt the case-cohort design because its simple weighting structure allows for straightforward extensions.

Despite their wide use, traditional analyses of case-cohort designs are statistically inefficient since they do not fully utilize covariates collected in phase 1, the full cohort. Previous studies have addressed this issue by treating unobserved covariate values as missing. In this setup, missingness depends only on case status and therefore satisfies the missing at random (MAR) assumption. Methods such as expectation–maximization (EM) algorithms and weighting-based approaches have been developed in this context. Although EM methods can yield efficiency gains under case-cohort sampling (\citealp{scheike2004maximum, nan2004efficient}), they are computationally demanding, particularly when the proportion of missing covariates is large (\citealp{kulathinal2007case}). Among weighting approaches, \citet{kulich2004improving} proposed doubly weighted estimators, while \citet{breslow2009improved, breslow2009using} applied the calibration framework of \citet{deville1992calibration} to improve the efficiency of hazard ratio estimation.

Another strategy for improving efficiency through the use of phase 1 covariates is multiple imputation (MI), originally proposed by \citet{rubin1987multiple}. Under MI, the incomplete dataset is imputed multiple times, typically 5 to 10 times. Each completed dataset is then analyzed separately, and the resulting estimates are combined using Rubin’s rules. While case-cohort analysis uses only the phase 2 sample, MI allows one to analyze the data as if the full cohort were observed by imputing values for units outside the phase 2 sample. In addition, MI avoids the intensive integration required in the E-step of EM-based methods and is straightforward to implement (\citealp{van2011mice}). Despite its strengths, standard MI can yield biased estimates when the analysis model includes nonlinear or interaction terms. Moreover, when the missing proportion of expensive biomarkers is high, the computational burden of MI grows rapidly as the full cohort size increases. These limitations highlight the need for MI procedures that remain unbiased in the presence of nonlinear or interaction terms and that are scalable while preserving statistical efficiency.

To address bias that arises when the analysis model is complex, MI has been discussed in terms of \textit{compatibility} and \textit{congeniality}, which describe the relationship between the imputation and analysis models. A set of conditional densities is said to be \textit{compatible} if there exists a joint distribution whose conditionals are identical to the set of conditional densities. While \citet{hughes2014joint} and \citet{liu2014stationary} referred to compatible conditional models as imputation models, \citet{bartlett2015multiple} viewed imputation and analysis models as such conditionals. A broader concept that appears in the literature is \textit{congeniality} studied by \citet{meng1994multiple} and \citet{xie2017dissecting}. In \citet{bartlett2015multiple}, compatibility is closely related to but weaker than congeniality, because compatibility does not require imputation from a well-defined Bayesian joint model. \citet{bartlett2015multiple} also proposed the substantive model compatible fully conditional specification (SMC-FCS) method, which incorporates a rejection sampling step into MI to ensure compatibility between the imputation and analysis models. By adopting SMC-FCS, we can mitigate the bias introduced by nonlinear or interaction terms in the analysis model.

Several works have studied MI in case-cohort and nested case-control designs using Cox proportional hazards regression as the analysis model (\citealp{white2009imputing, marti2011multiple, keogh2013using, keogh2018multiple}). In particular, \citet{keogh2013using} investigated an MI approach incorporating rejection sampling for these designs. By rejecting imputed values that are incompatible with the analysis model, SMC-FCS mitigates the bias introduced by interaction terms, as demonstrated in their simulation studies.

In large case–cohort studies, the computational burden becomes more pronounced for MI with rejection sampling because expensive biomarkers are observed only for a small subcohort. In a review of 32 published case-cohort analyses, \citet{sharp2014review} reported a median subcohort sampling fraction of 4.1\%. With such high levels of missingness, the computational cost of MI grows rapidly as the cohort size increases. To alleviate this burden, the \textit{supersampled case-cohort design} was recently proposed (\citealp{borgan2023use}). Rather than imputing the entire cohort outside the phase 2 sample, MI is applied only to a subset of individuals with missing data. Although MI on the supersample becomes scalable by lowering computational cost, it comes at the price of reduced efficiency of the estimator of interest. This is because additional units are selected without considering their contribution to the target estimator. The efficiency loss is particularly problematic in practice because the imputation procedure is typically performed only once on the observed dataset, unlike in simulation studies, where deviations from the true parameter value are averaged out over replications.

The influence function (IF) measures the first-order effect of each observation on an estimator and is commonly used to detect influential points in regression as well as to study the asymptotic properties of asymptotically linear estimators (\citealp{tsiatis2006semiparametric}). Observations with large IF magnitudes have a greater impact on the estimator. As a result, using IF to guide sampling allows influential units to be selected more frequently, improving efficiency (\citealp{ting2018optimal}).

In this paper, we propose an influence function-based supersampling (ISS) method for case-cohort studies. ISS is combined with weight calibration to reconcile two distinct sampling designs: case-cohort sampling and supersampling. The ISS procedure consists of two components. First, probability proportional to size (PPS) sampling (\citealp{hansen1943theory}) is employed, where the size measure is the $\ell_2$-norm of each unit’s influence function vector. Second, balanced sampling (\citealp{deville2004efficient}) is applied to obtain a supersample that satisfies balancing equations on the influence functions. Together, these steps allow sampling of units that are most informative for estimating the target parameter, such as the hazard ratio, which is the primary focus of our study. The proposed method achieves markedly higher efficiency in estimating log hazard ratios compared with RSS, without introducing bias. Compared with imputing the entire cohort, ISS substantially reduces the computational burden while preserving statistical efficiency.

In Section~\ref{sec:background}, we review traditional case–cohort designs and supersampling techniques that extend them. Section~\ref{sec:methods} presents our main methodology, including influence function–based supersampling, weight calibration, and the application of SMC-FCS within our framework. Section~\ref{sec:simulations} reports the simulation results, and Section~\ref{sec:casestudy} shows how the proposed method is applied to the National Institutes of Health--American Association of Retired Persons (NIH--AARP) Diet and Health Study. Finally, Section~\ref{sec:discussion} concludes with a discussion.

\section{Background}
\label{sec:background}
\subsection{Setup and Notation}
The parameter of interest is the log hazard ratio, $(\bm\beta_{\bm Z},\bm \beta_{\bm X})$, in the Cox proportional hazards model, 
$\lambda(t) = \lambda_0(t)\exp\!\left( \bm{\beta}_{\bm{Z}}^\top \bm{Z} + \bm{\beta}_{\bm{X}}^\top \bm{X} \right)$, possibly including interaction terms,
where $\lambda_0(t)$ denotes the baseline hazard function, $\bm{Z}=(Z_1,\ldots,Z_q)$ represents low-cost covariates, and $\bm{X}=(X_1,\ldots,X_p)$ represents expensive covariates. We denote the cumulative baseline hazard function by 
$\Lambda_0(t) = \int_0^t \lambda_0(u)\,du$. Low-cost covariates are variables that are relatively inexpensive to measure, such as demographic and behavioral variables (e.g., age, sex, smoking status). Expensive covariates, in contrast, are costly to obtain, such as blood biomarker assays or genomic data. For subject $i$, let $T_i$ and $C_i$ be the time-to-event and the censoring time, respectively. Then, we observe $\tilde T_i=\operatorname{min}(T_i, C_i)$ and $\delta_i=\mathbb{I}(T_i \le C_i)$, which denote the observed survival time and event indicator, respectively. The at-risk indicator at time $t$ is defined as $Y_i(t) = \mathbb{I}(\tilde T_i \ge t)$, and the counting process that jumps at the event time is $N_i(t) = \mathbb{I}(\tilde T_i \le t, \, \delta_i = 1)$.

In case-cohort studies, the full cohort is denoted by $\Omega$ with size $N$, and the randomly selected subset from the full cohort is the \textit{subcohort}, denoted by $\mathcal{SC}$ with size $n_{sc}$. The set of all cases is denoted by $\mathcal{D}$ with size $D$, and the number of cases in the subcohort is $d$. Let $m=n_{sc}-d$ denote the number of non-cases in the subcohort. By combining the subcohort $\mathcal{SC}$ and the set of all cases $\mathcal{D}$, we obtain the phase-2 sample, referred to as the \textit{case-cohort sample} denoted by $\mathcal{CC}$ with size $n_0 = n_{sc} + D$. Let $\Omega_{\mathrm{pool}}$ be the cohort outside the case-cohort sample. For supersampled case-cohort studies, we refer to the additional sample selected from $\Omega_{\mathrm{pool}}$ as the \textit{supersample}, denoted by $\mathcal{SS}$, with size $n_1$. Note that the supersample contains missing values in the expensive covariates $\bm X$. The union of the case-cohort sample and the supersample is referred to as the \textit{super case-cohort sample} with size $n=n_0+n_1$.

The empirical influence function of the $i^{\text{th}}$ unit for supersampling is denoted by $\hat{\bm{\psi}}_i=(\hat \psi_{i1},\ldots, \hat{\psi}_{iq})$. In computing the influence function values, we use the fully observed low-cost covariates, $\bm{Z}\in\mathbb{R}^{N\times q}$. We refer to the Cox proportional hazards model fitted with low-cost covariates as the \textit{submodel}. The model that includes both the low-cost covariates $\bm{Z}$ and the expensive covariates $\bm{X}$ is referred to as the \textit{full model}.

The missing at random (MAR) assumption holds when the probability of missingness depends only on the observed data, not on unobserved information. The MAR assumption is essential because standard MI is valid when data are missing at random or missing completely at random (MCAR). In case-cohort studies, selection of the phase 2 sample depends only on the survival status ($\delta_i$) and, when the stratified case-cohort design is employed, on the stratifying variables. Therefore, the MAR assumption holds, as phase 2 sampling relies solely on fully observed variables.

\subsection{Analysis of Case-Cohort Studies}
\label{sec:cc}
\subsubsection{Case-Cohort Studies}
For the case-cohort design, we estimate the log hazard ratio in the Cox proportional hazards model by maximizing the weighted pseudo-likelihood, $\mathcal{L}(\bm\beta)$. \citet{prentice1986case} used the term pseudo-likelihood to differentiate from the partial likelihood whose denominator is $\sum_{j=1}^N Y_j(t)exp(\bm\beta^\top \bm Z_j)$. The following equation estimates log hazard ratio for the case-cohort sample:

\begin{equation}
\bm{\hat{\beta}} 
= \argmax_{\bm\beta} \mathcal{L}(\beta) 
= \argmax_{\bm\beta} 
\prod_{i=1}^{n_0} \prod_{t > 0} 
\left\{ 
\frac{\exp\!\left(\bm \beta^{\top} \bm Z_i \right)}
{\sum_{j \in R(t)} w_j Y_j(t) \exp\!\left(\bm \beta^{\top} \bm Z_j \right)}
\right\}^{dN_i(t)},
\label{eq:pseudo_lik}
\end{equation}
where the weights $w_j$ are time-invariant, and the covariates are assumed to be measured at baseline for simplicity.

There are several estimation strategies for case-cohort studies. \citet{prentice1986case} set $w_j=1$ and $R(t)=\{i\in \mathcal{SC} \mid Y_i(t)=1\} \cup \{i \in \mathcal{D} \mid dN_i(t)=1\}$ in \eqref{eq:pseudo_lik}, thereby making cases outside the subcohort contribute only at their event time. \citet{self1988asymptotic} also suggested $w_j=1$ but made a slight change by setting $R(t)=\{i\in \mathcal{SC} \mid Y_i(t)=1\}$. The authors noted that the estimator is asymptotically equivalent to that of \citet{prentice1986case}, provided that the individual contribution to the score function is negligible. To enable cases outside the subcohort to contribute to the risk set at each event time, one may assign inverse-probability weights to $w_j$, following the approach of \citet{kalbfleisch1988likelihood}. Specifically,
\begin{equation}
    R(t) = \{i\in \mathcal{CC} \mid Y_i(t)=1\} \text{ and } w_j = \begin{cases} \dfrac{N-D}{m} & \text{if } j \in \mathcal{SC \setminus D} \\
    1 & \text{if } j \in \mathcal{D},
    \label{eq:cc_weight}
\end{cases}
\end{equation}
where $m=n_{sc}-d$ is the number of non-cases in the subcohort.

In this study, we adopt \eqref{eq:cc_weight} for case-cohort design due to its natural extension to supersampled case-cohort studies, as in \citet{borgan2023use}. Moreover, the weights defined in \eqref{eq:cc_weight} can be interpreted as a form of weight calibration, which allows a fair comparison between random supersampling and influence function-based supersampling in simulation studies. This will be further discussed in Section~\ref{sec:wc}. Following \citet{therneau1999computing}, we now refer to this estimation method under \eqref{eq:cc_weight} as \textit{Lin and Ying's estimate} (\citealp{lin1993cox}). 

The variance of the estimator in case-cohort studies consists of two components: (i) the standard Cox model variance (phase 1 variance) and (ii) the phase 2 variance. Following the implementation in the \texttt{survival} R package (Therneau et al., 2024), the variance of Lin and Ying’s estimator can be expressed as
\begin{equation}
    \mathcal{I}_{w}^{-1} + \left(1 - \frac{m}{N - D} \right) \sum_{i \in \mathcal{SC} \setminus \mathcal{D}}(\tilde{\bm{\psi}}_{i}-\bar{\bm{\psi}})(\tilde{\bm{\psi}}_{i}-\bar{\bm{\psi}})^\top,
    \label{eq:cc_var}
\end{equation}
where $\mathcal{I}_{w}$ denotes the observed information matrix from the full model fitted to the case-cohort sample, $\tilde {\bm{\psi}}_{i}$ is the corresponding influence function for observation $i$, and $\bar {\bm{\psi}} = \frac{1}{m} \sum_{\mathcal{SC}\setminus\mathcal{D}} \tilde {\bm{\psi}}_{i}$. For computing $\mathcal{I}_{w}$, the weights in \eqref{eq:cc_weight} are used. The first term in \eqref{eq:cc_var} corresponds to the phase 1 variance from the Cox model, and the second term represents the additional variability due to phase 2 sampling.

Lin and Ying's estimation strategy can be implemented by specifying \texttt{method = "LingYing"} of the \texttt{cch} function in the R \texttt{survival} package. Methods suggested in \citet{prentice1986case} and \citet{self1988asymptotic} can be implemented using \texttt{method = "Prentice"} and \texttt{method = "SelfPrentice"}, respectively.

\subsubsection{Supersampled Case-Cohort Studies}
\label{sec:cc_super}
In the random supersampled (RSS) case-cohort design, the risk set at time $t$ and the weights in \eqref{eq:pseudo_lik} are defined as
\begin{gather}
R(t) = \{i\in \mathcal{CC} \cup \mathcal{SS} \mid Y_i(t)=1\}, \label{eq:supercc_riskset} \\
w_j = \begin{cases} \dfrac{N-D}{m+n_1} & \text{if } j \in (\mathcal{SC \setminus D}) \cup \mathcal{SS} \\ 1 & \text{if } j \in \mathcal{D}, \label{eq:supercc_weight} \end{cases}
\end{gather}
by extending Lin and Ying's estimator. In the influence function-based supersampled (ISS) case-cohort design, the risk set remains the same, but calibrated weights are used instead of \eqref{eq:supercc_weight}. The calibration procedure is presented in Section~\ref{sec:wc}.

By treating the supersample as an increase in non-cases within the subcohort (\citealp{borgan2023use}), the variance for supersampled case-cohort studies then follows
\begin{equation}
    (\mathcal{I}_{w}^{S})^{-1}+\left(1 - \frac{m + n_1}{N - D} \right) \sum_{i \in (\mathcal{SC} \setminus \mathcal{D})\cup \mathcal{SS}}(\tilde{\bm{\psi}}_{i}^{S}-\bar{\bm{\psi}}^{S})(\tilde{\bm{\psi}}_{i}^{S}-\bar{\bm{\psi}}^{S})^\top,
    \label{eq:sscc_var}
\end{equation}
where $\mathcal{I}_{w}^{S}$ denotes the observed information matrix from the full model fitted to the super case-cohort sample, $\tilde {\bm{\psi}}_{i}^{S}$ is the corresponding influence function for observation $i$, and $\bar {\bm{\psi}}^{S} = \frac{1}{m+n_1} \sum_{(\mathcal{SC}\setminus\mathcal{D})\cup\mathcal{SS}} \tilde {\bm{\psi}}_{i}^{S}$. For computing $\mathcal{I}_{w}^{S}$, the weights are given by \eqref{eq:supercc_weight} under the RSS framework, and by the calibrated weights under the ISS framework.

\subsubsection{Stratified Supersampled Case-Cohort Studies}
In a stratified case-cohort design, the subcohort is randomly sampled within each stratum. \citet{borgan2000exposure} proposed three estimators for stratified case-cohort studies: Borgan I, II, and III. Among them, the Borgan II type estimator is a natural extension of Lin and Ying's estimator. It is asymptotically more efficient than the other two and allows for simpler variance estimation (\citealp{samuelsen2007stratified}). For stratified supersampled case-cohort analysis, where the supersample is added within each stratum, the risk set at time $t$ is the same as in \eqref{eq:supercc_riskset}, but the weights are given by
\begin{equation}
    w_{j} = \begin{cases} \dfrac{N_h-D_h}{m_h+n_{1h}} & \text{if } j \in (\mathcal{SC}_h \setminus \mathcal{D}_h)\cup\mathcal{SS}_h, \quad \forall h = 1,\ldots,H \\
    1 & \text{if } j \in \mathcal{D},
    \end{cases}
\label{eq:strcc_weight}
\end{equation}
where subscript $h$ denotes the $h^{\text{th}}$ stratum among the $H$ strata.

For variance estimation, we adopt the variance formula proposed by \citet{samuelsen2007stratified}, who reformulated the variance of the Borgan II type estimator using the influence function. In our simulation results, \citet{samuelsen2007stratified}'s variance formula yielded better coverage. Following the same approach as in the supersampled case-cohort analysis, the variance formula for the stratified supersampled case-cohort analysis is given by
\begin{equation}
    (\mathcal{I}_{w}^{SS})^{-1}+\sum_{h=1}^{H}\frac{m_h+ n_{1h}}{m_h+n_{1h}-1}\left(1-\frac{m_h+n_{1h}}{N_h-D_h}\right)\sum_{i \in (\mathcal{SC}_h \setminus \mathcal{D}_h)\cup \mathcal{SS}_h}(\tilde{\bm\psi}_{i}^{SS} - \bar{\bm\psi}^{SS}_h)(\tilde{\bm\psi}_{i}^{SS} -\bar{\bm\psi}^{SS}_h)^\top,
\end{equation}
where $\mathcal{I}_{w}^{SS}$ denotes the observed information matrix from the full model fitted to the stratified super case-cohort sample, $\tilde {\bm{\psi}}_{i}^{SS}$ is the corresponding influence function for observation $i$, and $\bar{\bm\psi}^{SS}_h=(m_h+n_{1h})^{-1}\sum_{i \in (\mathcal{SC}_h \setminus \mathcal{D}_h)\cup \mathcal{SS}_h} \tilde{\bm \psi}_{i}^{SS}$. Similar to the supersampled case–cohort analysis, the weights used in $\mathcal{I}_{w}^{SS}$ are given by \eqref{eq:strcc_weight} under the RSS framework, and by the calibrated weights under the ISS framework.

\section{Methods}
\label{sec:methods}
\subsection{Overview}
\label{sec:overview}
Before introducing the methodology, we first illustrate how the entire algorithm operates. Step~I in Algorithm~\ref{alg:overall_flow} summarizes the ISS procedure with weight calibration. Steps~II and~III correspond to the multiple imputation procedure for survival data with covariates missing by design.

ISS is designed to reduce the variance of the log hazard ratio estimator $\hat{\bm\beta}$ through efficiency gains in both the sampling and analysis stages. In the sampling stage, we employ a probability proportional to size (PPS) sampling design combined with balanced sampling. First, the inclusion probability $\pi_i^*$, proportional to the $\ell_2$-norm of the influence function vector for unit~$i$, $\|\hat{\bm\psi}_i\|_2 = \sqrt{\sum_{k=1}^q \hat \psi_{ik}^2}$, is computed. Then, the supersample is drawn to satisfy the balancing equations where the balancing variables for unit $i$ are denoted by $\bm B_i = (\pi_i^*, \hat\psi_{i1}, \ldots, \hat\psi_{iq})$.

To reduce the variance of the log hazard ratio in the analysis stage, we calibrate inverse probability weights (IPWs) to account for distinct sampling schemes: case-cohort sampling and supersampling. Through weight calibration, we reconcile the design weights from these two sampling schemes, which in turn helps stabilize the estimator $\hat{\bm\beta}$. Here, $\bm V_i$ denotes the sampling indicator, $d(\cdot,\cdot)$ is the distance metric, and $\bm A_i$ represents the auxiliary variables computed from phase~1 \citep{deville1992calibration}.

In the multiple imputation stage, the imputation is conducted $M$ times, and we iterate $L$ times for each imputation. The imputation model parameters, $\theta$, are sampled from their posterior distribution fitted within the subcohort, following the recommendation of \citet{keogh2013using}. This approach provides unbiased estimates because the subcohort is a random sample of the full cohort. When SMC-FCS is used, a rejection sampling step is added to the imputation algorithm. Although SMC-FCS effectively mitigates the bias that arises when the analysis model includes nonlinear or interaction terms, it increases computation time. SMC-FCS with ISS alleviates much of this computational burden by restricting imputation to a small supersample. Under this approach, the cumulative baseline hazard function $\Lambda_0(t)$ needs to be estimated using the calibrated weights computed in Step~I. However, existing R packages such as \texttt{mice} and \texttt{smcfcs} do not currently account for these features. The modified versions of these packages are available at \href{https://github.com/kjooho/ISS\_MI}{https://github.com/kjooho/ISS\_MI}.

In the final analysis, the log hazard ratio is computed using the $M$ imputed datasets. For the pooled variance, we sum the between-imputation variance with finite sample correction and the within-imputation variance. When computing the within-imputation variance $\bm V^{(m)}$, the phase 2 variance is added to the model variance.

\begin{algorithm}[htbp]
\small
\caption{Multiple Imputation using Influence Function-based Supersample}
\label{alg:overall_flow}
\begin{algorithmic}[1]
\State \textbf{Input} Full cohort data
\State \textbf{Output} Pooled log relative hazard(${\hat{\bm\beta}}$) and its variance
\vspace{0.25ex}

\State \textbf{I. Influence Function-Based Supersampling and Weighting}
\State Fit Cox PH model $\lambda(t \mid \bm{Z}) = \lambda_0(t) \exp(\bm Z \bm{\beta})$
\State Compute $\|\hat{\bm\psi}_i\|_2=\sqrt{\sum_{k=1}^q \hat \psi_{ik}^2} \quad \forall i=1,\ldots, N$
\State Compute $\pi_i^*=\min \left\{\lambda\|\hat{\bm\psi}_i\|_2,\, 1 \right\}$ with $\lambda$ satisfying $\sum_{i\in \Omega_{\mathrm{pool}}} \pi_i=n_1$
\State Draw a supersample satisfying $\sum_{i\in \Omega_{\mathrm{pool}}}\frac{\bm B_i}{\pi_i^*}V_i=\sum_{i\in \Omega_{\mathrm{pool}}}\bm B_i$ where $V_i \sim Bernoulli(\pi_i^*)$
\State Set initial weights $w_i^{0}= \bm{1}\left(i\in \mathcal{D}\right) + \frac{N}{n_{sc}}\bm{1}\left(i\in \mathcal{SC} \setminus \mathcal{D} \right) + \frac{1}{\pi_i^*}\bm{1}\left(i\in \mathcal{SS}\right)$
\State Calibrate $w^*_i = \argmin_{w_i} \sum_{i=1}^N \bm V_i\,d(w_i,w_i^0)\quad
\text{s.t. }\;
\sum_{i=1}^N \bm V_i w_i \bm A_i = \sum_{i=1}^N \bm A_i$
\State
\State \textbf{II. Multiple Imputation}
\For{$m=1,\ldots,M$}
\For{$k=1,\ldots,p$}
  \State Initialize $\bm X_{i}^{(0)}=\left(X_{i1}^{(0)},\ldots,X_{ip}^{(0)}\right), \text{ } \forall i \in \mathcal{SS}$.
  \For{$l=1,\ldots,L$}
    \State Sample $\theta_{X_k}^{(l)} \sim \pi(\theta\mid \bm X_{i}^{(l-1)}, \bm Z_i, \delta_i,\tilde T_i;\, i\in \mathcal{SC})$
    \State Sample $X_{ik}^{(l)}\sim f(X_{ik} \mid \bm X_{i,-k}^{(l-1)}, \bm Z_i, \delta_i,\tilde T_i; \theta_{X_k}^{(l)};\, i \in \mathcal{SS})$
    \If{\texttt{smcfcs}=True}
        \State Compute $\hat{\Lambda}_0(t) = \sum_{u \le t} \frac{dN(u)}{\sum_{r=1}^n Y_r(u)\, w_r \exp(\bm Z_{i} \bm{\hat{\beta}_Z} + \hat\beta_{X_k}^{\;\top} X_{rk}^{(l)} + \bm X_{r,-k}^{(l-1)} \bm{\hat{\beta}}_{\bm X_{\!-k}})}$
        \State Sample $U\sim \mathrm{Unif}(0,1)$
        \While{$U \leq \exp \{-\hat \Lambda_0(t) e^{\bm Z_{i} \bm{\hat{\beta}_Z} + \hat\beta_{X_k}^{\;\top} X_{rk}^{(l)} + \bm X_{r,-k}^{(l-1)} \bm{\hat{\beta}}_{\bm X_{\!-k}}}\}, \text{ } \forall i \in \mathcal{SS}$} 
            \State Repeat line $15-20$
        \EndWhile
    \EndIf
  \EndFor
\EndFor
\EndFor
\State
\State \textbf{III. Final Analysis}
\State Compute $\hat{\bm\beta}^{(m)} = \argmax_{\beta} \prod_{i=1}^{n} \prod_{t > 0} \left\{\frac{\exp\!\left( \bm Z_{i} \bm{\hat{\beta}_Z} + \bm X_{i}^{(L)} \bm{\hat{\beta}_{X}} \right)}{\sum_{r \in R(t)} w_r Y_r(t) \exp\!\left( \bm Z_{r} \bm{\hat{\beta}_Z} + \bm X_{r}^{(L)} \bm{\hat{\beta}_{X}} \right)} \right\}^{dN_i(t)}, \text{ } \forall m$
\State Compute within-variance $\bm V^{(m)}$ by adding phase 2 variance to the model variance
\State Return using Rubin's rule:
$\hat{\bm\beta}=\tfrac1M\!\sum_m\hat{\bm\beta}^{(m)},\;
 \operatorname{Var}(\hat{\bm\beta})=\tfrac1M\!\sum_m\bm V^{(m)}+(\frac{M+1}{M})\tfrac1{M-1}\!\sum_m(\hat{\bm\beta}^{(m)}-\hat{\bm\beta})^2$

\end{algorithmic}
\end{algorithm}

\subsection{Missing Scenarios and Supersampling}
\label{sec:scenarios}

\begin{figure}[htbp]
    \centering
    \begin{subfigure}{0.2512\textwidth}
        \centering
        \includegraphics[width=\textwidth]{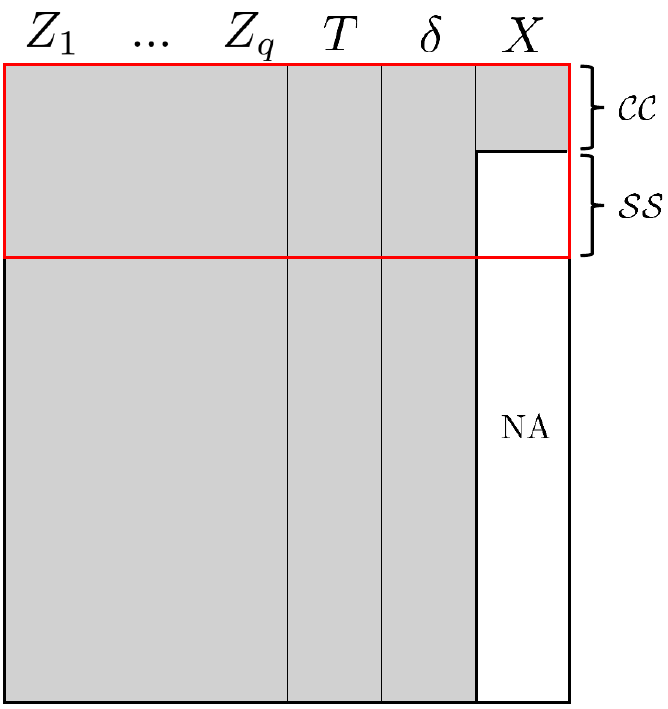}
        \caption{Single Missing}
        \label{fig:s1}
    \end{subfigure}
    \hfill
    \begin{subfigure}{0.32449\textwidth}
        \centering
        \includegraphics[width=\textwidth]{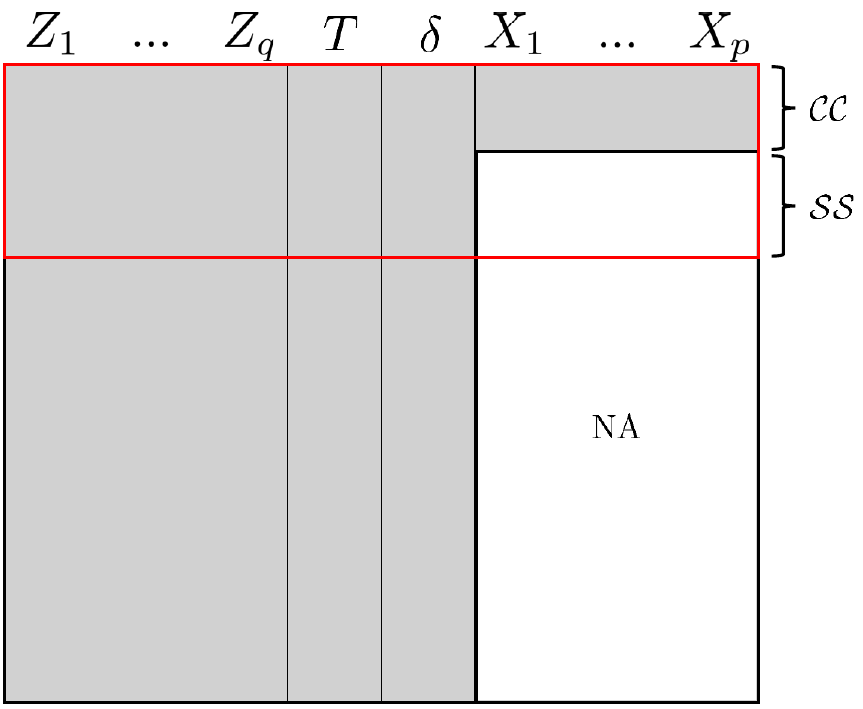}
        \caption{Within Study}
        \label{fig:s2}
    \end{subfigure}
    \hfill
    \begin{subfigure}{0.32449\textwidth}
        \centering
        \includegraphics[width=\textwidth]{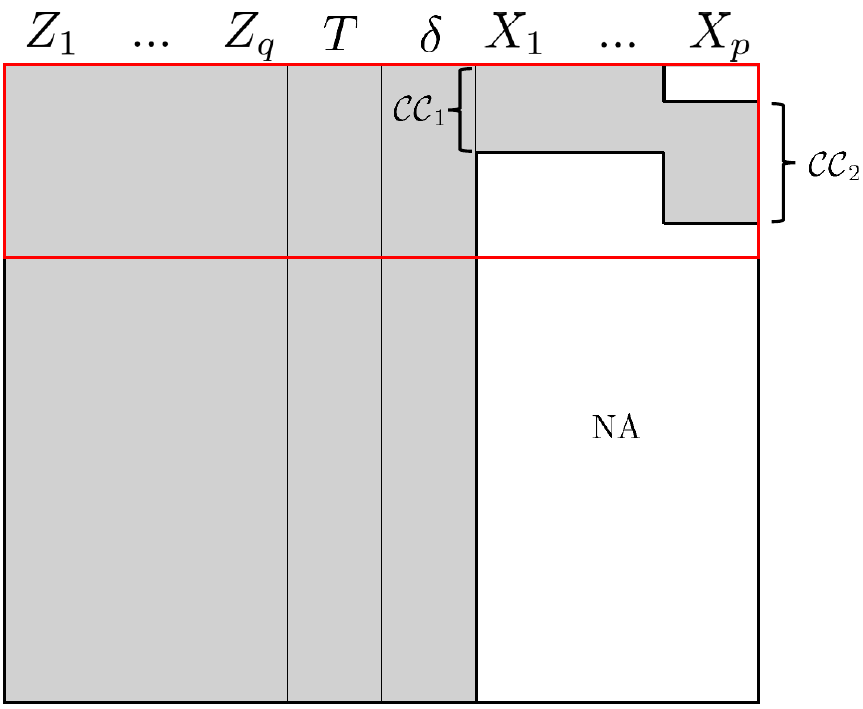}
        \caption{Between Studies}
        \label{fig:s3}
    \end{subfigure}
    \caption{\small The figures illustrate three types of missing data scenarios, where (a) shows a single expensive covariate with missingness, while (b) and (c) show multiple expensive covariates with missingness. Grey cells indicate observed data, and white cells with ``NA'' indicate missing values. The sample where $X$ is observed corresponds to the case-cohort sample, $\mathcal{CC}$. The red box indicates a super case-cohort sample, which consists of $\mathcal{CC}$ and $\mathcal{SS}$. In (c), $\mathcal{CC}_1$ denotes the case-cohort sample from one study and $\mathcal{CC}_2$ from another, and the white cells inside the red box represent the supersample for each study.}
\end{figure}

\subsubsection{Scenario 1: Single Expensive Covariate}
We now explore scenarios where the supersampling approach is most appropriate. The first scenario considers the case where the investigator is interested in assessing the effect of a single expensive biomarker $X$ on a specific event outcome $(\tilde T,\delta)$ using a case-cohort sample, as shown in Figure~\ref{fig:s1}. For instance, \citet{folsom2002c} investigated the association between C-reactive protein and coronary heart disease using data from the Atherosclerosis Risk in Communities (ARIC) Study. It is important to note that when $X$ is the only variable with missing values, more than one iteration in \texttt{mice} is unnecessary (\citealp{beesley2016multiple}). For \texttt{smcfcs}, when $X$ is missing due to the case-cohort design, the imputation model is fitted within the subcohort. In this setting, the imputed values and the estimated baseline cumulative hazard function do not depend on previous iterations, allowing the number of iterations to be reduced to one. As a result, imputing the full cohort using standard MI is already computationally efficient. Even so, MI combined with ISS is faster and achieves nearly the same efficiency as full-cohort imputation, even with a small supersample.

\subsubsection{Scenario 2: Multi-Dimensional Expensive Covariates}

In many epidemiological studies, a single blood assay or a sequencing sample provides measurements for many biomarkers. The second scenario considers multiple expensive covariates $\bm{X}\in \mathbb{R}^{N\times p}$ whose effects on the time-to-event outcome $(\tilde T, \delta)$ are of primary interest. In the ARIC Study, \citet{oluleye2013troponin} evaluated the associations of high-sensitivity troponin T (Hs-TnT), N-terminal pro--B-type natriuretic peptide (NT-proBNP), and high-sensitivity C-reactive protein (Hs-CRP) with mortality from various causes. More recently, \citet{deshotels2024vital} examined the associations between vital exhaustion and the same expensive biomarkers used in \citet{oluleye2013troponin}. Figure~\ref{fig:s2} illustrates such studies where many expensive biomarkers are collected within a study. In this scenario, increasing the number of iterations is crucial, as it allows imputation models to better capture the dependence structure of the data and achieve a stable set of imputations. However, in large cohorts with many expensive biomarkers, imputing all missing values can be computationally intensive, motivating the need for a supersampling framework. In such cases, ISS can facilitate the analysis while maintaining high efficiency.

Figure~\ref{fig:s3} depicts a scenario in which a researcher aims to combine distinct case-cohort samples from different studies. As the overlap between subcohorts may be limited, imputation can increase the efficiency of the estimator by leveraging available data. Recently, \citet{meng2025oral} integrated two prospective cohorts, the American Cancer Society Cancer Prevention Study-II Nutrition Cohort (CPS-II) and the Prostate, Lung, Colorectal, and Ovarian Cancer Screening Trial (PLCO), to assess the risk factors for pancreatic cancer. As more data from large-scale cohorts continue to be integrated, our proposed method will become increasingly useful in such scenarios. 

Another advantage of the supersampling framework arises when only a single variable $X_1$ exhibits missingness inside the red box of Figure~\ref{fig:s3}. In this setting, $\bm{X}_{-1}=(X_2,\ldots,X_p)$ are fully observed in the super case-cohort sample but missing outside it. If supersampling is adopted, this setting can be viewed as a special case of Scenario~1, where a single iteration of multiple imputation is sufficient. In contrast, imputing the remaining full cohort demands many iterations, thereby substantially increasing the computational burden.

\subsection{Influence Function-based Supersampling}
\label{sec:IF based SS}

\subsubsection{Influence Function of Hazard Ratio}
\label{sec:IF}
We denote $\hat{\bm{\psi}}_i$ as the influence function vector for $i^{\text{th}}$ observation from the submodel, and $\tilde{\bm{\psi}}_i$ as that from the full model. Recall that the influence function for supersampling, $\hat{\bm{\psi}}=(\hat{\bm{\psi}}_1,\ldots,\hat{\bm{\psi}}_N)^\top \in \mathbb{R}^{N\times q}$, is computed over fully observed variables $\bm Z \in \mathbb{R}^{N \times q}$, whereas $\tilde{\bm{\psi}}$ is based on $\bm Z$ and $\bm X$ for the case-cohort sample $\in \mathbb{R}^{n_0 \times (p+q)}$ or the super case-cohort sample $\in \mathbb{R}^{n \times (p+q)}$.

\citet{reid1985influence} introduced the influence function for the Cox proportional hazards regression, and \citet{chu2024} provided a detailed derivation in his doctoral dissertation. Here, we rewrite the empirical influence function for the hazard ratio in our notation. The empirical influence function $\hat{\bm{\psi}}_i \in \mathbb{R}^q$ for unit $i$ is
\begin{align}
\label{if_formula}
\hat{\bm{\psi}}_i = &\mathcal I^{-1}(\bm{\hat{\beta}})\, \delta_i\left\{\bm Z_i -\frac{\bm S^{(1)}(\bm{\hat\beta}, t)}{S^{(0)}(\bm{\hat\beta}, t)}\right\} \notag \\
& + \mathcal I^{-1}(\bm{\hat{\beta}})\exp(\bm{\hat\beta}^{\top}\bm Z_i) \left\{\sum_{t_j\le t_i}\delta_j \frac{\bm S^{(1)}(\bm{\hat\beta}, t)}{\{S^{(0)}(\bm{\hat\beta}, t)\}^2} - \bm Z_i \sum_{t_j\le t_i}\frac{\delta_j}{S^{(0)}(\bm{\hat\beta}, t)} \right\}.
\end{align}
where
\begin{gather*}
S^{(0)}(\bm \beta, t) = \sum_{k=1}^N Y_k(t)e^{\bm\beta^\top \bm Z_k} \text{ and }
\bm S^{(l)}(\bm \beta, t) = \sum_{k=1}^N Y_k(t)\bm Z_k^{\otimes l} e^{\bm\beta^\top \bm Z_k},\text{ } \forall l=1,2 , \\
\mathcal I(\bm{\beta}) = \sum_{i=1}^N \int_0^\tau \Bigg[ \frac{\bm S^{(2)}(\bm \beta, t)}{S^{(0)}(\bm \beta, t)} - \frac{\bm S^{(1)}(\bm \beta, t)\bm S^{(1)}(\bm \beta, t)^\top}{\{S^{(0)}(\bm \beta, t)\}^2} \Bigg]d N_i(t),
\end{gather*}
and $\mathcal I(\bm{{\beta}})$ indicates the observed information matrix, which is minus the second derivative of the log partial likelihood. The first term in \eqref{if_formula} measures the influence of the $i^{\text{th}}$ observation on the risk set at $t_i$ when unit $i$ is the case. The second term represents the influence of the $i^{\text{th}}$ observation on the risk set at $t_j \le t_i$ when unit $j$ is the case. Therefore, observations in the supersample have an influence function formula without the first term in \eqref{if_formula} as
\begin{equation}
    \hat{\bm{\psi}}_i = \mathcal I^{-1}(\bm {\hat{\beta}}) \exp(\bm{\hat\beta}^{\top}\bm Z_i)\Bigg\{
\sum_{t_j\le t_i}\delta_j \frac{\bm S^{(1)}(\bm{\hat\beta}, t)}
{\{S^{(0)}(\bm{\hat\beta}, t)\}^2} - \bm Z_i \sum_{t_j\le t_i}\frac{\delta_j}{S^{(0)}(\bm{\hat\beta}, t)}\Bigg\}.
\end{equation}
This is because cases who experience the event are included with probability one in case-cohort studies. Thus, the cohort from which the supersample is obtained only consists of non-cases, $\delta_i=0$, deleting the first term in \eqref{if_formula}. Based on this influence function formula, the next chapter demonstrates how ISS is conducted.

\subsubsection{Probability Proportional to Size (PPS) Sampling Design}
For ISS framework, we leverage probability proportional to size (PPS) sampling (\citealp{hansen1943theory}) and balanced sampling (\citealp{deville2004efficient}) to minimize the variance of the hazard ratio. We first explain why PPS sampling is adequate in our context. As shown in \citet{tsiatis2006semiparametric}, log hazard ratio $\hat{\bm\beta}$ is asymptotically linear having 
\begin{equation}
    \hat{\bm\beta} - \bm\beta = \frac{1}{N} \sum_{i=1}^{N} \bm\psi_i + o_p(N^{-1/2}),
\end{equation} 
where $\bm\beta$ and $\bm\psi_i$ are the true log hazard ratio and the true influence function. For sampling which units to impute, we may estimate $\sum_{i=1}^{N} \bm\psi_i$ by the Horvitz–Thompson (HT) estimator, $\sum_{i=1}^N \frac{V_i}{\pi_i}\hat{\bm\psi}_i$, where $V_i$ is the sampling indicator following Bernoulli distribution with probability $\pi_i$. In survey sampling, HT estimator provides an unbiased estimate for the population total of an auxiliary variable. Since $\hat{\bm\beta} - \bm\beta \approx \frac{1}{N} \sum_{i=1}^N \frac{V_i}{\pi_i}\hat{\bm\psi}_i$, the variance of $\hat{\bm\beta}$ can be approximated by
\begin{equation}
    \operatorname{Var}(\hat{\bm \beta}) \approx \frac{1}{N^2}\sum_{i=1}^{N} \frac{(1 - \pi_i)}{\pi_i} \hat{\bm\psi}_i\hat{\bm\psi}_i^\top + \frac{2}{N^2}\sum_{i=1}^{N} \sum_{j>i} \frac{(\pi_{ij} - \pi_i \pi_j)}{\pi_i \pi_j} \hat{\bm\psi}_i \hat{\bm\psi}_j^\top ,
    \label{eq:HTvar}
\end{equation}
using the variance formula for HT estimator (\citealp{cochran1977sampling}). We aim to find the optimal inclusion probabilities that minimize the first term of \eqref{eq:HTvar}. Although minimizing \eqref{eq:HTvar} including the second term is possible, doing so substantially increases the computation time for obtaining the inclusion probabilities, thereby undermining the main advantage of the supersampling framework. Moreover, in large cohorts where supersampling is advantageous, the joint inclusion probability can be asymptotically approximated by $\pi_{ij} \approx \pi_i \pi_j$, making the second term negligible.

Following the approach of \citet{ting2018optimal}, who suggested minimizing the trace of the variance when estimating multiple parameters, the optimal inclusion probabilities $\pi_i^*$ are obtained by
\begin{equation}
    \pi_i^{*}= \underset{{\pi_i}} \argmin \sum_{i\in \Omega_{\mathrm{pool}}}\frac{(1 - \pi_i)}{\pi_i} \|\hat{\bm\psi}_i\|_2^2\,\, \text{ subject to } \sum_{i\in \Omega_{\mathrm{pool}}} \pi_i=n_1,
\end{equation}
which simplifies to
\begin{equation}
    \pi_i^{*}=\min \left\{\lambda \|\hat{\bm\psi}_i\|_2,\, 1\right\}\,\, \text{ subject to } \sum_{i\in \Omega_{\mathrm{pool}}} \pi_i=n_1.
\end{equation}
This corresponds to a PPS sampling design, where the size measure is defined as the $\ell_2$-norm of the influence function vector for unit $i$, $\|\hat{\bm\psi}_i\|_2$.

For the stratified case–cohort design, the optimal inclusion probabilities are obtained within each stratum by solving
\begin{equation}
    \pi_i^{*}=\min \left\{\lambda \|\hat{\bm\psi}_i\|_2,\, 1\right\}\,\, \text{ subject to } \sum_{i\in \Omega_{\mathrm{pool},h}} \pi_i=n_{1h}, \quad \forall h=1,\ldots,H,
    \label{eq:strcc_incl_prob}
\end{equation}
where $\sum_{h=1}^H n_{1h}=n_1$ and $\sum_{h=1}^H\Omega_{\mathrm{pool},h}=\Omega_{\mathrm{pool}}$. 

\subsubsection{Balanced Sampling}
Balanced sampling is well aligned with the PPS design, as it builds upon the obtained optimal inclusion probabilities. Incorporating balanced sampling into the PPS framework further reduces the variance of the HT estimator. This is because the variance depends only on the residuals from the regression of the target variable on the balancing variables (\citealp{tille2011ten}). Furthermore, balanced sampling mitigates the occurrence of negative or extreme weights that may arise when calibration is applied alone (\citealp{deville2004efficient}). The efficiency gain from balanced sampling becomes more pronounced as the correlation between the balancing variables and the target parameter increases.

Following the definition of balanced sampling in \citet{deville2004efficient}, a sampling design is said to be balanced on the auxiliary variables $\bm x_{1},\ldots \bm x_q$ if and only if it satisfies the balancing equations given by
\begin{equation}
    \sum_{i\in \Omega} \frac{V_i}{\pi_i}x_{ij} = \sum_{i \in \Omega}{x_{ij}} \quad \forall j =1,\ldots,k,
    \label{eq:balacingeq}
\end{equation}
where $V_i \sim Bernoulli (\pi_i)$ and $\pi_i$ is the inclusion probability. However, in practice, the balancing equations cannot always be exactly satisfied. Hence, a computational procedure that approximates the equations while maintaining the prescribed inclusion probabilities is needed. The \textit{cube method} provides such procedure. Based on \eqref{eq:balacingeq}, the cube method can be regarded as finding a sampling indicator vector $\bm V = (V_1, \ldots, V_{N})$ in the intersection of the $N$-cube, $[0,1]^N$, and the affine subspace $Q\in\mathbb{R}^{N-k}$ where 
\begin{equation}
  Q = \Bigl\{\,\bm{V}\in\mathbb{R}^{N} \;\Big|\;
       \sum_{i\in \Omega} \frac{V_i}{\pi_i}\bm{x}_i = \sum_{i \in \Omega}{{\bm{x}}_i}
    \Bigr\}.
\end{equation}

In case-cohort studies using ISS, we draw a balanced sample of size $n_1$ based on the optimal inclusion probabilities $\pi_i^*$ from the PPS design. The cube method finds a sampling indicator vector $\tilde{\bm V} = (V_1, \ldots, V_{N-n_0})$ in the intersection of the $(N-n_0)$-cube, $[0,1]^{N-n_0}$, and the affine subspace $\tilde Q\in\mathbb{R}^{(N-n_0)-(q+1)}$ where 
\begin{equation}
  \tilde Q = \Bigl\{\,\tilde{\bm V}\in\mathbb{R}^{N-n_0} \;\Big|\;
       \sum_{i\in \Omega_{\mathrm{pool}}} \frac{V_i}{\pi_i^*}\bm{B}_i = \sum_{i \in \Omega_{\mathrm{pool}}}{\bm{B}_i}
    \Bigr\},
    \label{eq:affine_subsp}
\end{equation}
with $V_i \sim Bernoulli(\pi_i^*)$ and $\bm B_i = (\pi_i^*, \hat\psi_{i1}, \ldots, \hat\psi_{iq})$. To ensure a fixed-size balanced sample, we incorporate the inclusion probability $\pi_i^*$ into $\bm B_i$ since $\sum_{i\in \Omega_{\mathrm{pool}}}\pi_i^*=n_1$. 

In stratified case-cohort studies, the supersample is selected within each stratum with probabilities proportional to $\|\hat{\bm\psi}_i\|_2$, ensuring that the balancing equations are satisfied. Therefore, the balancing equations in \eqref{eq:affine_subsp} are modified as
\begin{equation}
    \sum_{i\in \Omega_{\mathrm{pool},h}} \frac{V_i}{\pi_i^*}\bm{B}_i = \sum_{i \in \Omega_{\mathrm{pool},h}}{\bm{B}_i},
\end{equation}
where $\pi_i^*$ is defined in \eqref{eq:strcc_incl_prob}.

In our study, the target parameter is $\frac{1}{N} \sum_{i=1}^{N} \bm\psi_i$ where $\bm \psi_i$ is the true influence function from the full model. While it is ideal to use the empirical influence function $\tilde{\bm{\psi}}_i$ from the full model for all $i\in\Omega_{\mathrm{pool}}$, this is infeasible because the covariates $\bm X$ are only partially observed. Although $\tilde{\bm{\psi}}_i$ can be computed after a single imputation of $\bm X$, this approach tends to increase the estimator’s empirical variance due to the additional noise introduced by single imputation. Consequently, satisfying the balancing equations in \eqref{eq:affine_subsp} with $\hat{\bm{\psi}}_i$ minimizes the design variance of the log hazard ratio to the greatest extent achievable.

\subsection{Weight Calibration}
\label{sec:wc}
\subsubsection{Generalized Raking}
While PPS design with balanced sampling reduces the design variance in the sampling stage, weight calibration can reduce the variance in the analysis stage. Weight calibration is a method that uses auxiliary information of the phase 1 sample to improve the efficiency of the estimator of interest.

Let $\bm A_i$ denote auxiliary variables for unit $i$ that are computed from phase 1 data. Given $\bm A_i$, the calibrated weights are given as
\begin{equation}
    w_i^*=\argmin_{w_i} \sum_{i\in \Omega} \bm V_i \, d(w_i, w_i^0)\,\, \text{ subject to } \sum_{i\in \Omega} \bm V_i w_i \bm A_i \;=\; \sum_{i\in \Omega} \bm A_i
\end{equation}
where $d(\cdot, \cdot)$ is a distance measure and $\bm V_i$ is the sampling indicator. Because linear distance, $d(a,b)=(a-b)^2/2b$, may yield negative weights, we choose $d(a,b)=a\log(\frac{a}{b})+b-a$ (\citealp{deville1992calibration}, Case 2) which is referred to as \textit{raking} or \textit{exponential tilting}. Similar to balanced sampling, a stronger association between the auxiliary variables $\bm A_i$ and the target parameter leads to a more efficient estimator.

\subsubsection{Supersampled Case-Cohort Analysis with Calibrated Weights}
In our proposed calibration equations, weight calibration not only improves the efficiency of log hazard ratio, but also reconciles the two distinct sampling designs, the case-cohort sampling and supersampling. For supersampled case-cohort studies, natural calibration equations with $\bm A_i=\bm (1,1)^\top$ would be 
\begin{gather}
    \sum_{i\in \Omega \setminus \mathcal{D}} I\left(i \in (\mathcal{SC \setminus D}) \cup \mathcal{SS}\right) w_i = N-D ,\label{eq:wc_rss1}\\ 
    \sum_{i\in \mathcal{D}} I\left(i \in \mathcal{D}\right)w_i = D.
    \label{eq:wc_rss2}
\end{gather}
These are precisely the calibration equations implicit in random supersampled case-cohort studies (\citealp{borgan2023use}) where weights are defined as $w_i=\frac{N-D}{m+n_1}$ if $i \in (\mathcal{SC \setminus D}) \cup \mathcal{SS}$, and $w_i=1$ if $i \in \mathcal{D}$.

However, in the ISS framework, the inclusion probabilities for the non-case subcohort and those for the supersample are derived from distinct sampling designs. Specifically, the non-case subcohort is selected as a simple random sample from the full cohort $\Omega$, whereas the supersample is drawn from the remaining cohort $\Omega_{\mathrm{pool}}$ using the PPS design that satisfies the balancing equations. Therefore, the indicator $V_i = I\left(i \in (\mathcal{SC \setminus D}) \cup \mathcal{SS}\right)$ in \eqref{eq:wc_rss1} must be decomposed into two separate indicators, $I(i \in \mathcal{SC \setminus D})$ and $I(i \in \mathcal{SS})$.

The main challenge lies in determining how to allocate the sample size $N-D$ between the two non-overlapping samples, as no strict criterion exists for this partition. To address this, we split the sample size $N-D$ in proportion to the sum of the $\ell_2$-norm of the influence function vectors within each sample. That is, the calibration equations are given as
\begin{gather}
    \sum_{i\in \Omega \setminus \mathcal{D}} I\left(i \in \mathcal{SC \setminus D}\right)w_i = (N-D)\frac{db_{0}}{db_{0}+db_{1}}, \label{eq:wc_iss1} \\
    \sum_{i\in \Omega \setminus \mathcal{D}} I(i \in \mathcal{SS})w_i = (N-D)\frac{db_{1}}{db_{0}+db_{1}}, \label{eq:wc_iss2} \\
    \sum_{i\in \mathcal{D}} I(i \in \mathcal{D})w_i = D,
    \label{eq:wc_iss3}
\end{gather}
where $db_0=\sum_{i \in \mathcal{SC \setminus D}} \|\hat{\bm{\psi}}_i\|_2$ and $db_1=\sum_{i \in \mathcal{SS}} \|\hat{\bm{\psi}}_i\|_2$. The sum of \eqref{eq:wc_iss1} and \eqref{eq:wc_iss2} exactly reproduces \eqref{eq:wc_rss1} since the non-case subcohort and the supersample are disjoint. In this way, samples with greater overall influence receive larger weights, aligning with the size measure defined in the sampling stage.

In a stratified case-cohort design, calibrating weights within each stratum substantially increases the number of calibration variables. This leads to greater variability in the weights and, consequently, to a larger sample variance (\citealp{breslow2009improved}). Therefore, instead of calibrating within each stratum, we use the same calibration equations as those in the case-cohort analysis (\eqref{eq:wc_iss1}, \eqref{eq:wc_iss2}, \eqref{eq:wc_iss3}).

\subsection{Substantive Model Compatible Fully Conditional Specification}
\label{sec:smcfcs}
SMC-FCS, proposed by \citet{bartlett2015multiple}, helps eliminate bias when the analysis model includes interaction or nonlinear effects of covariates. First, we can construct an imputation model that is compatible with the analysis model,
\begin{equation}
    f(X_{k} \mid \bm X_{-k}, \bm Z, \delta, \tilde T) \propto f(\tilde T, \delta \mid \bm X, \bm Z, \bm \beta) f(X_{k} \mid \bm X_{-k}, \bm Z, \theta_{X_k}),
\end{equation}
where $f(\tilde T, \delta \mid \bm X, \bm Z, \bm \beta)$ is the analysis model, with $\bm \beta$ representing its parameter, and $\theta_{X_k}$ indicating the imputation model parameter. We refer to this imputation model $f(X_{k} \mid \bm X_{-k}, \bm Z, \delta, \tilde T)$ as the \textit{target density}. Although drawing samples from the target density is ideal, this is often intractable. Therefore, we sample from a \textit{proposal density} $f(X_{k} \mid \bm X_{-k}, \bm Z, \theta_{X_k})$, which is easier to sample from, and reject the imputed values that are far from the target density. Specifically, the ratio of the target density to the proposal density is bounded above by some function $c(\tilde T, \delta,\bm X_{-k}, \bm Z, \theta_{X_k})$ which does not contain $X_k$. Then, we accept the imputed values $X_k^*$ from the proposal density if 
\begin{equation}
    U \le \frac{f(\tilde T, \delta \mid X_k^*, \bm X_{-k}, \bm Z, \bm \beta)}{c(\tilde T, \delta,\bm X_{-k}, \bm Z, \theta_{X_k})}
\end{equation}
where $U \sim \mathrm{Unif}(0,1)$. MI with rejection sampling ensures compatibility with the analysis model and mitigates the bias in MICE that arises in the presence of nonlinear or interaction effects. Note that, unlike MICE, the imputation predictors should be independent of the outcome, given the predictors in the analysis model.

Implementing SMC-FCS under a supersampled case-cohort design requires several modifications to the standard SMC-FCS (\citealp{keogh2013using}). First, the conditional model for $X_{k} \mid \bm X_{-k}, \bm Z, \delta, \tilde T, \theta_{X_k}$ is fitted using the subcohort. Next, imputation is performed only for the missing values in the supersample rather than for the entire cohort. In the rejection sampling step, the weighted baseline cumulative hazard function must be used because the analysis is conducted on the super case-cohort sample. That is,
\begin{equation}
    \hat{\Lambda}_0(t) = \sum_{u \le t} \frac{dN(u)}{\sum_{r=1}^n Y_r(u)\, w_r \exp(\bm Z_{i} \bm{\hat{\beta}_Z} + \hat\beta_{X_k}^{\;\top} X_{rk}^{(l)} + \bm X_{r,-k}^{(l-1)} \bm{\hat{\beta}}_{\bm X_{\!-k}})}.
    \label{eq:smc_cumbhaz}
\end{equation}
Accordingly, the weights $w_r$ in \eqref{eq:smc_cumbhaz} are the calibrated weights from the ISS framework described in Section~\ref{sec:wc}. Then the imputed values are accepted if
\begin{equation}
    U \leq \exp \{-\hat \Lambda_0(t) e^{\bm Z_{i} \bm{\hat{\beta}_Z} + \hat\beta_{X_k}^{\;\top} X_{rk}^{(l)} + \bm X_{r,-k}^{(l-1)} \bm{\hat{\beta}}_{\bm X_{\!-k}}}\}, \text{ } \forall i \in \mathcal{SS},
\end{equation}
assuming $T \perp C \mid \bm X, \bm Z$ and $T \perp \bm X \mid \bm Z$.

\section{Simulation Studies}
\label{sec:simulations}
\subsection{Setup}
We generated a total of ten covariates: four low-cost, fully observed covariates and six expensive covariates subject to missingness under the case-cohort design. The low-cost covariates include two continuous variables $(\mathrm{z}_0, \mathrm{z}_1)$, one binary variable $\mathrm{z}_2$, and one categorical variable $\mathrm{z}_3$ with three levels. The dummy variables corresponding to $\mathrm{z}_3$ are denoted by $\mathrm{z}_{3.2}=\mathbb{I}(\mathrm{z}_3=2)$ and $\mathrm{z}_{3.3}=\mathbb{I}(\mathrm{z}_3=3)$. The continuous variables $(\mathrm{z}_0, \mathrm{z}_1)$ follow a bivariate normal distribution $N_2(\mathbf{0}, \Sigma_2)$, where the diagonal elements of $\Sigma_2$ are one and $\text{corr}(\mathrm{z}_0, \mathrm{z}_1)=0.05$. The binary covariate $\mathrm{z}_2$ is obtained by thresholding a latent standard normal variable $\mathrm{u}_2$ correlated with $\mathrm{z}_0$ and $\mathrm{z}_1$, ensuring $P(\mathrm{z}_2=1)=0.5$, $\text{corr}(\mathrm{z}_0,\mathrm{z}_2)=-0.05$, and $\text{corr}(\mathrm{z}_1,\mathrm{z}_2)=0.01$. The categorical variable $\mathrm{z}_3$ is generated from a multinomial logistic model:
$$
\mathbb{P}(\mathrm{z}_{3}=j)=\frac{e^{\eta_{j}}}{\sum_{k=1}^3 e^{\eta_{k}}}, \text{ where }
\begin{pmatrix}
\eta_{1}  \\
\eta_{2} \\
\eta_{3}
\end{pmatrix} 
= 
\begin{pmatrix}
0 & 0 \\
-0.5 & -0.1 \\
-0.3 & -0.2 
\end{pmatrix} 
\begin{pmatrix}
\mathrm{z}_0  \\
\mathrm{z}_1
\end{pmatrix}.
$$
The six expensive covariates consist of four continuous variables $(\mathrm{xc}_1,\ldots,\mathrm{xc}_4)$ and two binary variables $(\mathrm{xb}_1, \mathrm{xb}_2)$. The continuous expensive covariates follow a linear model,
$$
\begin{pmatrix}
\mathrm{xc}_1  \\
\mathrm{xc}_2 \\
\mathrm{xc}_3 \\
\mathrm{xc}_4
\end{pmatrix} 
= 
\begin{pmatrix}
0.2 & 0.1 & 0.1 & 0.1 & -0.1 \\
0.1 & -0.15 & 0.1 & 0.1 & 0.05 \\
0.05 & -0.1 & 0.15 & -0.05 & 0.1 \\
0.2 & 0.01 & -0.1 & 0.12 & -0.05
\end{pmatrix} 
\begin{pmatrix}
\mathrm{z}_0  \\
\mathrm{z}_1 \\
\mathrm{u}_2 \\
\mathrm{z}_{3.2} \\
\mathrm{z}_{3.3}
\end{pmatrix}
 + \bm\varepsilon
$$
where $\bm\varepsilon \sim N(\bm 0, I_4)$.
The binary expensive covariates are generated from a logistic model:
$$
P(\mathrm{xb}_j = 1) = 
\frac{1}{1+e^{-\tilde{\eta}_j}}, \text{ where }
\begin{pmatrix}
\tilde{\eta}_{1}  \\
\tilde{\eta}_{2}
\end{pmatrix} 
= 
\begin{pmatrix}
0.15 & 0.1 & 0.07 & 0.08 & -0.03 \\
0.15 & 0.15 & 0 & 0.15 & -0.05
\end{pmatrix} 
\begin{pmatrix}
\mathrm{z}_0  \\
\mathrm{z}_1 \\
\mathrm{u}_2 \\
\mathrm{z}_{3.2} \\
\mathrm{z}_{3.3}
\end{pmatrix}.
$$

The proportional hazards model is generated from a Weibull$(\gamma, \alpha)$ distribution so that the survival function is $S_i(t\,; \bm Z_i, \bm X_i) = \exp\{-(\gamma_i t)^{\alpha}\}$ with $\gamma^{\alpha}_i = \exp(\bm Z_i \bm\beta_{\bm Z} + \bm X_i \bm\beta_{\bm X} + \beta_{\mathrm{z1:xc_1}}^\top \mathrm{z}_{1}\mathrm{xc}_{1}),\,\, \forall i=1,\ldots,N$. 
The hazard function is 
$$
\lambda_i(t\,; \bm Z_i, \bm X_i) = \lambda_0(t) \exp(\bm Z_i \bm\beta_{\bm Z} + \bm X_i \bm\beta_{\bm X}+\beta_{\mathrm{z1:xc_1}}^\top \mathrm{z}_{1}\mathrm{xc}_{1}),
$$
where $\lambda_0(t) = \alpha t^{\alpha - 1} \exp(\beta_0)$, $\bm Z_i=(\mathrm{z}_{i1}, \mathrm{z}_{i2}, \mathrm{z}_{i3.2}, \mathrm{z}_{i3.3})$, and $\bm X_i=(\mathrm{xc}_{i1},\ldots, \mathrm{xc}_{i4}, \mathrm{xb}_{i1}, \mathrm{xb}_{i2})$. If an interaction term between $\mathrm{z_1}$ and $\mathrm{xc_1}$ is present in the analysis model, the true log hazard ratios are given by
\begin{align*}
    &(\beta_{\mathrm{z_1}}, \beta_{\mathrm{z_2}}, \beta_{\mathrm{z_{3.2}}}, \beta_{\mathrm{z_{3.3}}}, \beta_{\mathrm{xc_1}}, \beta_{\mathrm{xc_2}}, \beta_{\mathrm{xc_3}}, \beta_{\mathrm{xc_4}}, \beta_{\mathrm{xb_1}}, \beta_{\mathrm{xb_2}}, \beta_{\mathrm{z1:xc_1}})^\top \\
    &= (1.5, 0.5, 0.1, 0.2, 0.4, 0.1, 0.1, 0.1, 0.3, 0.5, 0.3)^\top.
\end{align*}
Without the interaction term, $\beta_{\mathrm{z1:xc_1}}=0$. All subjects enter randomly within the first two years, $T_0 \sim U(0, 2)$, and the administrative censoring time is 15 years after the study begins. We define $C_1 = 15 - T_0$ as the time from entry to administrative censoring. We also consider censoring from death due to other causes, $C_2 \sim \text{Weibull}(-\log(0.9)/15, 1)$, for 10\% death at $t = 15$. We define the event time as the minimum of survival and censoring times, $\tilde{T} = \min(T, C_1, C_2)$, and a case indicator for the event of interest, $\delta = I\{T \leq \min(C_1, C_2)\}$. We consider the time-on-study time scale.

Following the simulation settings described in \citet{borgan2023use}, we conducted 1,000 simulation replicates, each based on a full cohort of 25,000 individuals, a subcohort of 250 individuals, and approximately 250 cases. For the supersampling design, 750 additional units were selected for imputation, corresponding to the small supersample in \citet{borgan2023use}. For all methods, the number of iterations was fixed at 20, as the computational burden increases substantially when multiple expensive covariates are included in the analysis model. Moreover, following the recommendation of \citet{white2009imputing}, the Nelson Aalen estimator $\hat \Lambda_0(\tilde T)$, rather than the observed survival time $\tilde T$, was included in the imputation models for MICE. In SMC-FCS, as mentioned in Section~\ref{sec:smcfcs}, only the low-cost covariates $\bm Z_i$ are included as predictors of the imputation model.

\subsection{Simulation Results}
In Table~\ref{tab:sim1}, case-cohort analysis shows bias higher than that of all six imputation methods, with relative efficiency around 10\% and coverage below 90\% across all covariates. The coverage is particularly low for $\mathrm{z_1}$ (52.4\%), which has a strong effect, with a true hazard ratio of $\exp(1.5)$. These results highlight the need for imputation methods to achieve unbiased and efficient estimation.

Without an interaction term in the analysis model, bias across all imputation methods tends to be mild. Still, the RSS methods exhibit greater bias toward fully observed covariates with stronger effects, including $\mathrm{z_1}$ and $\mathrm{z_2}$. This is consistent with simulation studies by \citet{amorim2021two}, who showed that random sampling tends to yield higher bias when the true log hazard ratio, $\beta$, deviates from the null. 

As expected, the ISS methods demonstrate high relative efficiency for low-cost covariates, comparable to that of imputing the entire cohort (MICE, SMC). The differences in relative efficiency between the ISS and imputing the whole cohort are below 5\% for all low-cost covariates. In contrast, the RSS methods are less than half as efficient as the ISS methods for low-cost covariates. For expensive covariates, the efficiency gain of ISS over RSS is modest but consistently positive.

The empirical coverage appears to be around 95\% for all imputation methods except for MICE RSS and SMC RSS methods in $\mathrm{z_1}$, which are 84.1\% and 81.7\%, respectively.

The computation time (in seconds) of the supersampled methods is significantly lower than that of imputing the entire cohort, yielding approximately a 95\% reduction in runtime for SMC with supersampling relative to the standard SMC method. The differences in computation time between the RSS and ISS methods are negligible. When the analysis model does not include any interaction terms, there is little advantage in using SMC-based methods over MICE-based methods. 

\begin{table}[htbp]
\centering
\begin{adjustbox}{max width=\textwidth}
\begin{threeparttable}
\captionsetup{font=large, labelfont=bf, labelsep=period}
\caption{Simulation Results: Analysis Model Without an Interaction Term}
\label{tab:sim1}
\begin{tabular}{lcccccccc}
\toprule
\textbf{Statistics} & \textbf{Full Cohort} & \textbf{Case-cohort} & \textbf{MICE} & \textbf{MICE RSS} & \textbf{MICE ISS} & \textbf{SMC} & \textbf{SMC RSS} & \textbf{SMC ISS} \\
\midrule
avg.time & 0.363 & 0.018 & 37.758 & 3.997 & 4.169 & 1767.819 & 71.308 & 81.388 \\
max time & 0.735 & 0.057 & 49.994 & 4.736 & 5.097 & 1871.735 & 81.960 & 100.521 \\
min time & 0.275 & 0.012 & 32.583 & 3.010 & 3.409 & 1578.670 & 37.555 & 37.022 \\
\specialrule{0.08em}{0.3em}{0.3em}
bias $\mathrm{z}_1$ & 0.004 & 0.252 & 0.031 & 0.112 & 0.025 & 0.053 & 0.122 & 0.048 \\
bias $\mathrm{z}_2$ & 0.002 & 0.058 & 0.005 & 0.030 & 0.008 & 0.014 & 0.030 & 0.017 \\
bias $\mathrm{z}_{3.2}$ & 0.002 & 0.004 & 0.001 & 0.006 & 0.001 & 0.002 & 0.003 & 0.0004 \\
bias $\mathrm{z}_{3.3}$ & 0.006 & 0.017 & 0.006 & 0.003 & 0.006 & 0.0005 & 0.010 & 0.0009 \\
bias $\mathrm{xc}_1$ & 0.001 & 0.060 & 0.029 & 0.001 & 0.021 & 0.007 & 0.014 & 0.001 \\
bias $\mathrm{xc}_2$ & 0.002 & 0.014 & 0.009 & 0.0002 & 0.007 & 0.003 & 0.003 & 0.003 \\
bias $\mathrm{xc}_3$ & 0.0002 & 0.007 & 0.012 & 0.004 & 0.009 & 0.007 & 0.004 & 0.005 \\
bias $\mathrm{xc}_4$ & 0.0006 & 0.011 & 0.002 & 0.000 & 0.003 & 0.002 & 0.005 & 0.0004 \\
bias $\mathrm{xb}_1$ & 0.001 & 0.038 & 0.011 & 0.013 & 0.023 & 0.024 & 0.028 & 0.024 \\
bias $\mathrm{xb}_2$ & 0.005 & 0.060 & 0.016 & 0.034 & 0.040 & 0.042 & 0.047 & 0.044 \\
\specialrule{0.08em}{0.3em}{0.3em}
mc.se $\mathrm{z}_1$ & 0.070 & 0.222(9.8\%) & 0.079(77.7\%) & 0.140(24.8\%) & 0.081(73.6\%) & 0.082(72.4\%) & 0.141(24.4\%) & 0.085(67.7\%) \\
mc.se $\mathrm{z}_2$ & 0.132 & 0.416(10.0\%) & 0.160(68.0\%) & 0.239(30.3\%) & 0.160(67.6\%) & 0.162(66.4\%) & 0.242(29.6\%) & 0.165(64.1\%) \\
mc.se $\mathrm{z}_{3.2}$ & 0.151 & 0.467(10.5\%) & 0.183(68.5\%) & 0.272(30.9\%) & 0.186(66.3\%) & 0.188(64.9\%) & 0.282(28.6\%) & 0.191(62.9\%) \\
mc.se $\mathrm{z}_{3.3}$ & 0.152 & 0.489(9.7\%) & 0.183(69.6\%) & 0.271(31.6\%) & 0.183(69.7\%) & 0.185(68.1\%) & 0.281(29.3\%) & 0.187(66.3\%) \\
mc.se $\mathrm{xc}_1$ & 0.064 & 0.197(10.7\%) & 0.130(24.6\%) & 0.141(21.1\%) & 0.132(24.0\%) & 0.138(21.8\%) & 0.146(19.4\%) & 0.140(21.2\%) \\
mc.se $\mathrm{xc}_2$ & 0.062 & 0.190(10.7\%) & 0.131(22.5\%) & 0.139(19.9\%) & 0.130(23.0\%) & 0.139(20.1\%) & 0.143(18.9\%) & 0.141(19.5\%) \\
mc.se $\mathrm{xc}_3$ & 0.064 & 0.200(10.3\%) & 0.140(21.0\%) & 0.151(17.9\%) & 0.142(20.3\%) & 0.147(18.9\%) & 0.155(17.1\%) & 0.149(18.5\%) \\
mc.se $\mathrm{xc}_4$ & 0.063 & 0.196(10.3\%) & 0.129(23.8\%) & 0.141(20.0\%) & 0.129(23.9\%) & 0.135(21.7\%) & 0.146(18.5\%) & 0.138(20.8\%) \\
mc.se $\mathrm{xb}_1$ & 0.127 & 0.393(10.5\%) & 0.269(22.4\%) & 0.285(19.9\%) & 0.253(25.3\%) & 0.279(20.8\%) & 0.294(18.7\%) & 0.285(20.0\%) \\
mc.se $\mathrm{xb}_2$ & 0.133 & 0.391(11.5\%) & 0.271(23.9\%) & 0.288(21.1\%) & 0.262(25.5\%) & 0.288(21.1\%) & 0.300(19.5\%) & 0.290(20.8\%) \\
\specialrule{0.08em}{0.3em}{0.3em}
est.se $\mathrm{z}_1$ & 0.068(0.945) & 0.138(0.524) & 0.082(0.954) & 0.130(0.841) & 0.093(0.971) & 0.083(0.920) & 0.128(0.817) & 0.094(0.953) \\
est.se $\mathrm{z}_2$ & 0.130(0.949) & 0.300(0.837) & 0.167(0.958) & 0.240(0.956) & 0.190(0.979) & 0.167(0.958) & 0.241(0.947) & 0.192(0.978) \\
est.se $\mathrm{z}_{3.2}$ & 0.150(0.954) & 0.355(0.875) & 0.190(0.962) & 0.279(0.958) & 0.216(0.976) & 0.190(0.959) & 0.280(0.946) & 0.218(0.981) \\
est.se $\mathrm{z}_{3.3}$ & 0.157(0.954) & 0.366(0.853) & 0.198(0.965) & 0.290(0.964) & 0.225(0.983) & 0.198(0.962) & 0.288(0.967) & 0.227(0.982) \\
est.se $\mathrm{xc}_1$ & 0.064(0.943) & 0.145(0.827) & 0.131(0.954) & 0.159(0.977) & 0.145(0.971) & 0.131(0.935) & 0.156(0.964) & 0.141(0.951) \\
est.se $\mathrm{xc}_2$ & 0.064(0.950) & 0.146(0.857) & 0.132(0.953) & 0.158(0.984) & 0.142(0.968) & 0.131(0.946) & 0.157(0.961) & 0.142(0.958) \\
est.se $\mathrm{xc}_3$ & 0.063(0.956) & 0.145(0.849) & 0.132(0.941) & 0.158(0.957) & 0.141(0.955) & 0.131(0.907) & 0.157(0.956) & 0.142(0.926) \\
est.se $\mathrm{xc}_4$ & 0.063(0.958) & 0.144(0.846) & 0.131(0.965) & 0.155(0.970) & 0.139(0.957) & 0.130(0.950) & 0.154(0.966) & 0.140(0.961) \\
est.se $\mathrm{xb}_1$ & 0.129(0.955) & 0.296(0.856) & 0.269(0.959) & 0.318(0.971) & 0.277(0.967) & 0.265(0.940) & 0.316(0.957) & 0.284(0.947) \\
est.se $\mathrm{xb}_2$ & 0.135(0.956) & 0.301(0.876) & 0.271(0.950) & 0.320(0.971) & 0.280(0.964) & 0.269(0.945) & 0.318(0.953) & 0.286(0.951) \\
\bottomrule
\end{tabular}
\caption*{The table summarizes simulation results for eight methods estimating log hazard ratios under the analysis model without an interaction term. The average, maximum, and minimum computation times for each iteration are reported. Bias is computed as the absolute bias, \textit{mc.se} denotes the Monte Carlo standard error, and \textit{est.se} indicates the estimated standard error obtained from the Cox model. The parentheses in \textit{mc.se} indicate the relative efficiency compared to the \textit{Full Cohort}, while the parentheses in \textit{est.se} show the empirical coverage. \textit{Full Cohort} applies the Cox model to the complete dataset without missingness, which is the gold standard. \textit{Case-cohort} fits the Cox model using only the case-cohort sample. \textit{MICE} and \textit{SMC} impute the full cohort outside the case-cohort sample using MICE and SMC-FCS algorithms. \textit{MICE RSS} and \textit{SMC RSS} apply the MICE and SMC-FCS algorithms to the random supersample, respectively. \textit{MICE ISS} and \textit{SMC ISS} apply them to the influence function-based supersample. \textit{MICE} and \textit{SMC} can be considered the gold-standard approaches for supersampling analyses.}
\end{threeparttable}
\end{adjustbox}
\end{table}

\begin{figure}[htbp]
    \centering
    \includegraphics[width=\textwidth]{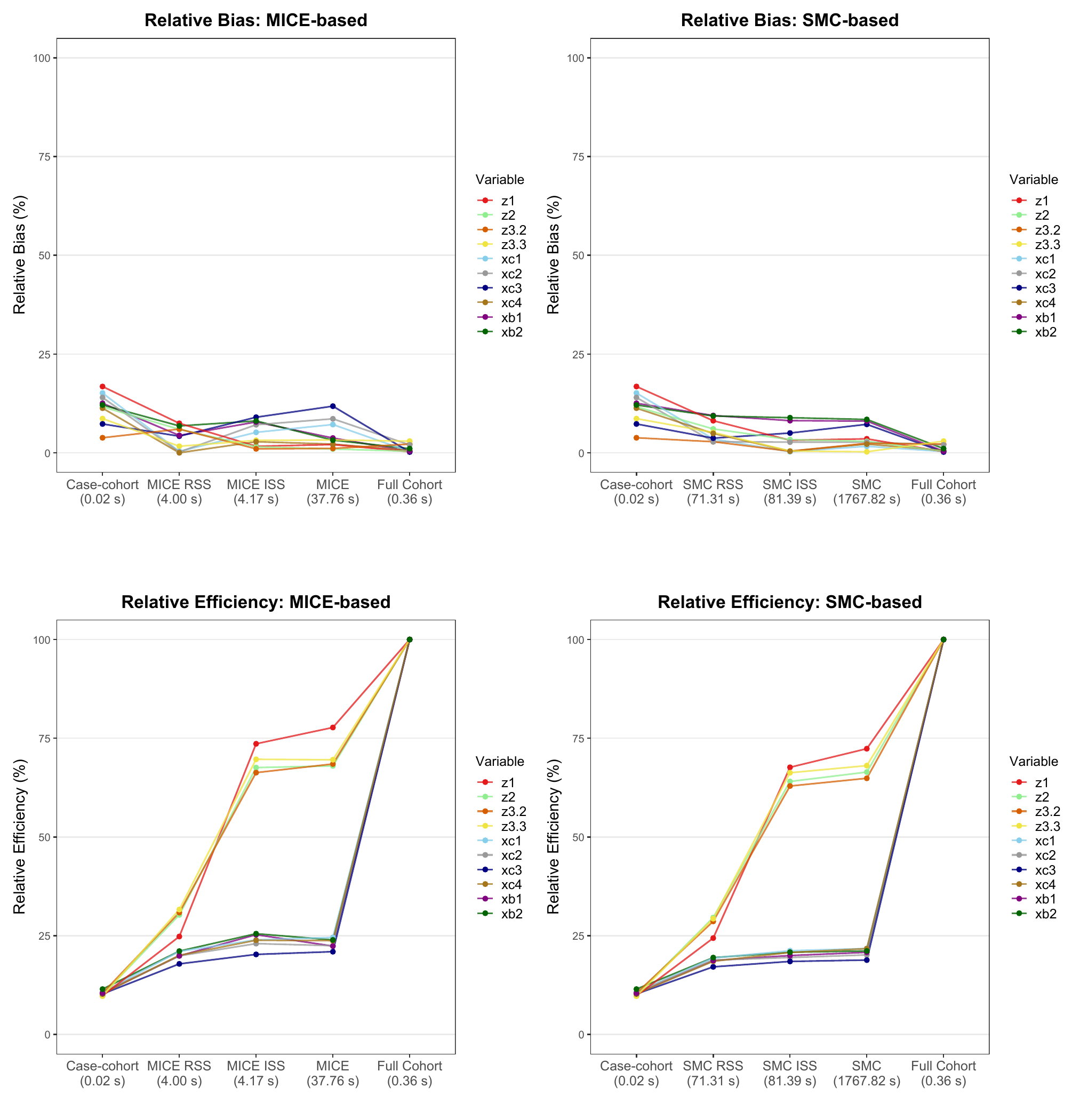}
    \captionsetup{font=small, labelfont=bf, labelsep=period}
    \caption{Relative Bias and Efficiency: Analysis Model Without an Interaction Term}
    \caption*{The top left and right panels show the relative bias of MICE-based and SMC-based methods, respectively, while the bottom left and right panels show their relative efficiency. The x-axis is ordered such that efficiency is expected to increase, reaching 100\% for the Full Cohort. The ISS methods achieve markedly higher efficiency than the RSS methods, with efficiency levels close to those of imputing the entire cohort. Numbers in parentheses below the x-axis indicate the runtime (in seconds) for each method. The SMC ISS method reduced the computation time by approximately 95\% compared to the SMC method. When the analysis model does not include an interaction term, bias remains small across all methods.}
    \label{fig:simulation1}
\end{figure}

\newpage
However, when an interaction term is present in the analysis model, as in Table~\ref{tab:sim2}, the MICE-based methods yield biased estimates for the interaction variable ($\mathrm{z}_1 \times \mathrm{xc}_1$) and the covariates involved ($\mathrm{z}_1$ and $\mathrm{xc}_1$). Consequently, the coverage for $\mathrm{z}_1 \times \mathrm{xc}_1$ is poor in the MICE and MICE ISS methods. The coverage of MICE RSS for the interaction term is relatively high, primarily because of its inflated estimated variance. In contrast, SMC-based methods exhibit much lower bias, particularly in $\mathrm{z}_1$ and $\mathrm{xc}_1$. These results are consistent with the literature comparing MICE and SMC-FCS (\citealp{keogh2013using, bartlett2015multiple, van2018flexible}).

Similar to the results in Table~\ref{tab:sim1}, the efficiency gain is substantial in ISS methods compared to the RSS methods in both MICE and SMC-based methods.

\begin{table}[htbp]
\centering
\begin{adjustbox}{max width=\textwidth}
\begin{threeparttable}
\captionsetup{font=large, labelfont=bf, labelsep=period}
\caption{Simulation Results: Analysis Model With an Interaction Term}
\label{tab:sim2}
\begin{tabular}{lcccccccc}
\toprule
\textbf{Statistics} & \textbf{Full Cohort} & \textbf{Case-cohort} & \textbf{MICE} & \textbf{MICE RSS} & \textbf{MICE ISS} & \textbf{SMC} & \textbf{SMC RSS} & \textbf{SMC ISS} \\
\specialrule{0.08em}{0.3em}{0.3em}
avg.time & 0.394 & 0.019 & 40.416 & 4.185 & 4.365 & 1964.690 & 77.002 & 99.144 \\
max time & 0.824 & 0.066 & 48.244 & 5.696 & 5.623 & 2092.693 & 86.429 & 181.538 \\
min time & 0.318 & 0.014 & 34.361 & 3.559 & 3.735 & 1764.143 & 40.646 & 42.600 \\
\specialrule{0.08em}{0.3em}{0.3em}
bias $\mathrm{z}_1$ & 0.003 & 0.245 & 0.084 & 0.143 & 0.075 & 0.067 & 0.126 & 0.061 \\
bias $\mathrm{z}_2$ & 0.004 & 0.076 & 0.029 & 0.008 & 0.025 & 0.007 & 0.034 & 0.011 \\
bias $\mathrm{z}_{3.2}$ & 0.004 & 0.002 & 0.005 & 0.006 & 0.004 & 0.0006 & 0.005 & 0.0009 \\
bias $\mathrm{z}_{3.3}$ & 0.002 & 0.025 & 0.021 & 0.004 & 0.017 & 0.0002 & 0.010 & 0.003 \\
bias $\mathrm{xc}_1$ & 0.002 & 0.140 & 0.159 & 0.092 & 0.180 & 0.044 & 0.051 & 0.065 \\
bias $\mathrm{xc}_2$ & 0.001 & 0.014 & 0.015 & 0.003 & 0.013 & 0.006 & 0.000 & 0.003 \\
bias $\mathrm{xc}_3$ & 0.0002 & 0.013 & 0.013 & 0.0002 & 0.009 & 0.003 & 0.0003 & 0.004 \\
bias $\mathrm{xc}_4$ & 0.001 & 0.012 & 0.008 & 0.0003 & 0.007 & 0.002 & 0.002 & 0.0009 \\
bias $\mathrm{xb}_1$ & 0.003 & 0.017 & 0.002 & 0.010 & 0.009 & 0.021 & 0.021 & 0.023 \\
bias $\mathrm{xb}_2$ & 0.013 & 0.075 & 0.005 & 0.034 & 0.025 & 0.045 & 0.053 & 0.052 \\
bias $\mathrm{z}_1\times \mathrm{xc}_1$ & 0.003 & 0.030 & 0.190 & 0.085 & 0.191 & 0.050 & 0.020 & 0.053 \\
\specialrule{0.08em}{0.3em}{0.3em}
mc.se $\mathrm{z}_1$ & 0.090 & 0.262(11.7\%) & 0.100(80.8\%) & 0.157(32.6\%) & 0.103(76.5\%) & 0.102(76.9\%) & 0.155(33.4\%) & 0.106(72.2\%) \\
mc.se $\mathrm{z}_2$ & 0.132 & 0.410(10.3\%) & 0.175(56.5\%) & 0.254(26.9\%) & 0.177(55.5\%) & 0.183(51.6\%) & 0.265(24.7\%) & 0.189(48.5\%) \\
mc.se $\mathrm{z}_{3.2}$ & 0.157 & 0.487(10.4\%) & 0.203(60.2\%) & 0.297(28.1\%) & 0.206(58.4\%) & 0.211(55.6\%) & 0.292(28.9\%) & 0.213(54.1\%) \\
mc.se $\mathrm{z}_{3.3}$ & 0.159 & 0.505(9.9\%) & 0.205(59.9\%) & 0.305(27.1\%) & 0.207(58.7\%) & 0.215(54.7\%) & 0.312(25.9\%) & 0.220(52.1\%) \\
mc.se $\mathrm{xc}_1$ & 0.125 & 0.274(20.7\%) & 0.154(65.7\%) & 0.184(45.8\%) & 0.164(58.0\%) & 0.164(57.8\%) & 0.189(43.5\%) & 0.177(49.3\%) \\
mc.se $\mathrm{xc}_2$ & 0.066 & 0.212(9.7\%) & 0.140(22.4\%) & 0.153(18.7\%) & 0.142(21.8\%) & 0.148(19.9\%) & 0.154(18.4\%) & 0.152(18.9\%) \\
mc.se $\mathrm{xc}_3$ & 0.066 & 0.204(10.3\%) & 0.139(22.1\%) & 0.149(19.3\%) & 0.140(21.8\%) & 0.149(19.3\%) & 0.156(17.7\%) & 0.153(18.4\%) \\
mc.se $\mathrm{xc}_4$ & 0.066 & 0.200(10.8\%) & 0.136(23.4\%) & 0.147(19.8\%) & 0.138(22.5\%) & 0.142(21.2\%) & 0.151(18.9\%) & 0.148(19.7\%) \\
mc.se $\mathrm{xb}_1$ & 0.132 & 0.414(10.2\%) & 0.281(22.0\%) & 0.303(19.0\%) & 0.269(24.0\%) & 0.297(19.8\%) & 0.314(17.7\%) & 0.304(18.9\%) \\
mc.se $\mathrm{xb}_2$ & 0.143 & 0.430(11.1\%) & 0.278(26.5\%) & 0.308(21.7\%) & 0.271(28.1\%) & 0.299(23.0\%) & 0.328(19.1\%) & 0.307(21.8\%) \\
mc.se $\mathrm{z}_1\times \mathrm{xc}_1$ & 0.070 & 0.171(16.9\%) & 0.076(84.5\%) & 0.100(49.2\%) & 0.081(75.2\%) & 0.085(67.4\%) & 0.103(45.9\%) & 0.091(59.5\%) \\
\specialrule{0.08em}{0.3em}{0.3em}
est.se $\mathrm{z}_1$ & 0.086(0.946) & 0.169(0.635) & 0.106(0.903) & 0.165(0.876) & 0.118(0.949) & 0.100(0.910) & 0.156(0.900) & 0.114(0.943) \\
est.se $\mathrm{z}_2$ & 0.129(0.946) & 0.292(0.830) & 0.184(0.961) & 0.281(0.969) & 0.213(0.980) & 0.180(0.943) & 0.273(0.963) & 0.212(0.975) \\
est.se $\mathrm{z}_{3.2}$ & 0.148(0.937) & 0.343(0.828) & 0.211(0.971) & 0.330(0.967) & 0.243(0.981) & 0.205(0.947) & 0.318(0.967) & 0.242(0.975) \\
est.se $\mathrm{z}_{3.3}$ & 0.158(0.946) & 0.360(0.838) & 0.221(0.959) & 0.342(0.968) & 0.255(0.989) & 0.214(0.945) & 0.331(0.959) & 0.255(0.981) \\
est.se $\mathrm{xc}_1$ & 0.122(0.937) & 0.213(0.852) & 0.167(0.902) & 0.227(0.974) & 0.213(0.956) & 0.166(0.950) & 0.213(0.971) & 0.211(0.977) \\
est.se $\mathrm{xc}_2$ & 0.063(0.933) & 0.141(0.798) & 0.141(0.954) & 0.173(0.970) & 0.151(0.967) & 0.133(0.925) & 0.165(0.961) & 0.144(0.935) \\
est.se $\mathrm{xc}_3$ & 0.063(0.938) & 0.140(0.826) & 0.139(0.955) & 0.174(0.975) & 0.153(0.971) & 0.132(0.921) & 0.164(0.961) & 0.145(0.942) \\
est.se $\mathrm{xc}_4$ & 0.063(0.945) & 0.139(0.827) & 0.136(0.944) & 0.172(0.975) & 0.150(0.969) & 0.131(0.931) & 0.163(0.964) & 0.143(0.943) \\
est.se $\mathrm{xb}_1$ & 0.129(0.937) & 0.291(0.828) & 0.284(0.960) & 0.353(0.979) & 0.298(0.968) & 0.270(0.934) & 0.336(0.964) & 0.293(0.931) \\
est.se $\mathrm{xb}_2$ & 0.135(0.938) & 0.299(0.824) & 0.285(0.957) & 0.354(0.977) & 0.300(0.974) & 0.273(0.921) & 0.341(0.956) & 0.295(0.950) \\
est.se $\mathrm{z}_1\times \mathrm{xc}_1$ & 0.069(0.940) & 0.119(0.843) & 0.098(0.548) & 0.160(0.988) & 0.117(0.700) & 0.085(0.927) & 0.131(0.978) & 0.106(0.970) \\
\bottomrule
\end{tabular}
\caption*{The table summarizes the simulation results for eight methods estimating log hazard ratios under an analysis model with an interaction term. MICE-based methods tend to show higher bias than SMC-based methods for both the interaction term and the covariates involved. As in Table~\ref{tab:sim1}, the efficiency differences between ISS methods and standard MI methods remain small, indicating that efficiency gains from ISS methods are largely preserved even when the analysis model includes interactions.}
\end{threeparttable}
\end{adjustbox}
\end{table}

\begin{figure}[htbp]
    \centering
    \includegraphics[width=\textwidth]{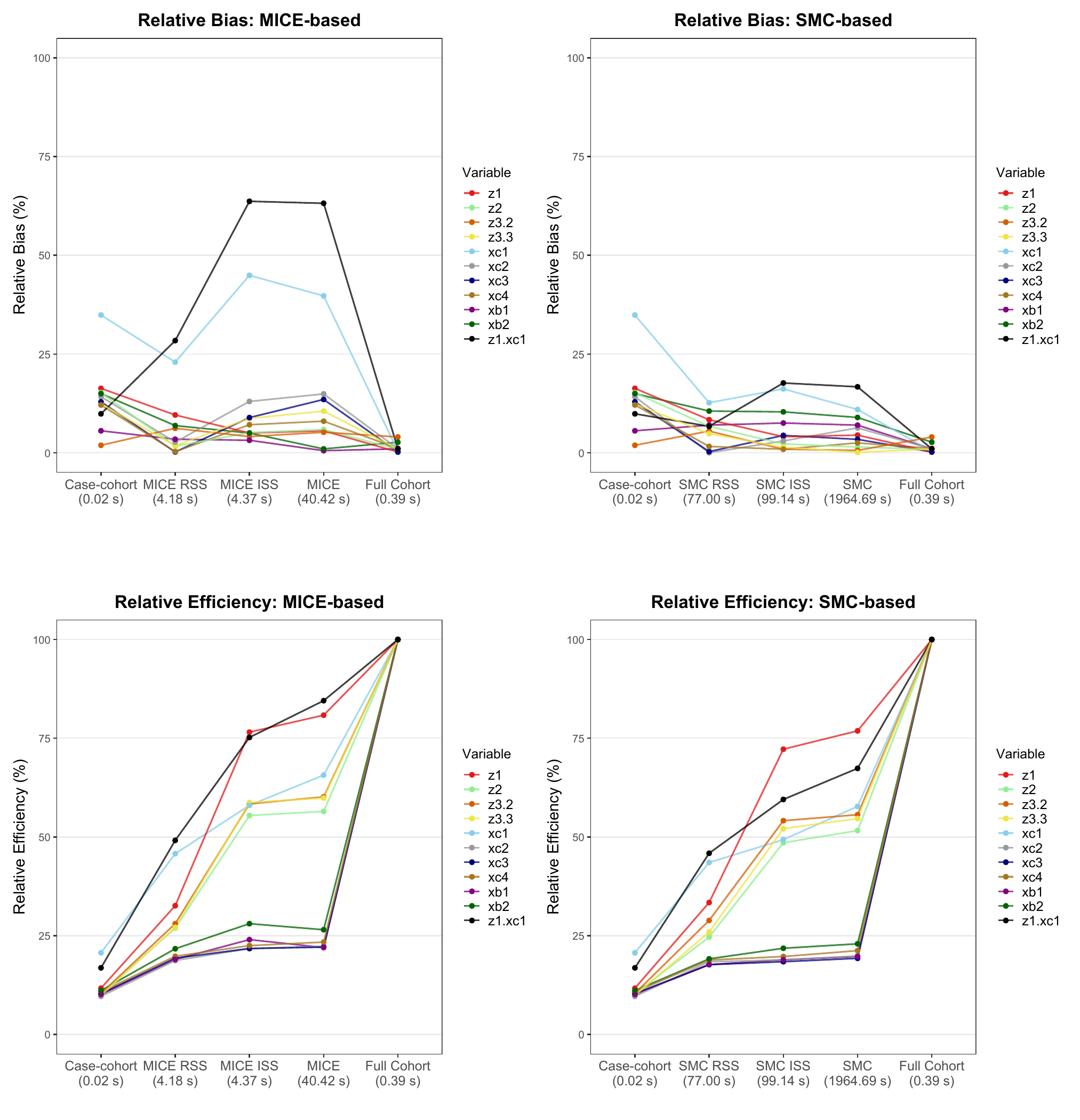}
    \captionsetup{font=small, labelfont=bf, labelsep=period}
    \caption{Relative Bias and Efficiency: Analysis Model With an Interaction Term}
    \caption*{As in Figure~\ref{fig:simulation1}, the top left and right panels show the relative bias of MICE-based and SMC-based methods, respectively, while the bottom left and right panels show their relative efficiency. The MICE-based methods exhibit substantial relative bias in the interaction term and its associated variables. In contrast, SMC-based imputation methods tend to yield small bias. The apparently high efficiency of MICE-based methods for the interaction term results from this bias rather than genuine precision gains. As illustrated in Figure~\ref{fig:simulation1}, ISS methods achieve efficiency comparable to that of imputing the entire cohort while requiring significantly lower computational time.}
    \label{fig:simulation2}
\end{figure}

\section{Case Study}
\label{sec:casestudy}
To illustrate our proposed methods, we analyzed data from the National Institutes of Health--American Association of Retired Persons (NIH--AARP) Diet and Health Study, which recruited participants aged 50-71 years between 1995 and 1996 \citep{schatzkin2001design}. To enable direct comparison with the full cohort analysis based on complete data, individuals with missing values in any predictors used in either the analysis or imputation model were excluded. Subsequently, multiple imputation and analysis were performed once for each method. Participants with implausible measurements--waist circumference outside the range of 20-60 inches and daily caloric intake outside 500-5,000 kcal--were also removed. Continuous variables, including waist circumference, caloric intake, and entry age, were standardized. For interpretability, waist circumference was standardized within sex. After data cleaning, the full cohort for this analysis included 232,453 participants, among whom 337 were diagnosed with pancreatic cancer during follow-up through December 2000. 

The low-cost covariates included sex (0: female, 1: male), smoking status (0: never, 1: quit $\le 20$ cigs/day, 2: quit $>20$ cigs/day, 3: current $\le 20$ cigs/day, 4: current $>20$ cigs/day), race (1: Non-Hispanic White, 2: Non-Hispanic Black, 3: Hispanic, 4: Asian/Other), age group (0: 50-59, 1: 60-64, 2: 65-71 years), and history of diabetes (0: no, 1: yes). In addition to these variables, the analysis model also included an interaction between sex and waist circumference, given previous evidence that the association between waist circumference and pancreatic cancer differs by sex (\citealp{stolzenberg2008adiposity}). The imputation model further incorporated physical activity (0: never, 1: rarely, 2: 1-3 times/month, 3: 1-2 times/week, 4: 3-4 times/week, 5: $\ge 5$ times/week) and continuous entry age. Waist circumference and daily caloric intake were treated as expensive covariates and set to be missing outside the stratified case-cohort sample. The stratified case-cohort sample consisted of 1,000 subcohort members and all 337 cases. The stratifying variables used to construct the phase 2 sample were sex and age group, forming six strata. The table summarizing the strata and the number of participants is presented in Appendix Table~\ref{tab:strata_summary}.

From the full-cohort estimates, we observe that current smoking and age over 65 substantially increase the hazard of being diagnosed with pancreatic cancer. In addition, the hazard for men is $\exp(0.236)=0.27$, that is, 27\% higher than that for women. A one-standard-deviation increase in waist circumference is associated with an approximately 12\% increase in hazard for men and a 4\% increase for women. Therefore, an increase in waist circumference is associated with a greater risk for men.

In Table~\ref{tab:casestudy}, compared with RSS methods, ISS methods show smaller bias from the full cohort estimates. For the interaction term ($\mathrm{sex \times waist}$), MICE-based methods show larger bias than those observed in the SMC-FCS methods. Although MICE ISS shows bias similar to that of the standard SMC method, this is likely due to the randomness arising from a single run. The number of iterations was set to 500 following the rule of thumb suggested by \citet{borgan2023use}, who reported that 200 iterations could lead to biased estimates in SMC-FCS. The runtime is notably long for the standard SMC-FCS, taking approximately 7.7 hours, whereas the supersampling versions of SMC-FCS require only about 8 minutes.

\begin{table}[htbp]
\centering
\begin{adjustbox}{max width=\textwidth}
\begin{threeparttable}
\captionsetup{font=large, labelfont=bf, labelsep=period}
\caption{Comparison of methods using the NIH--AARP Diet and Health Study}
\label{tab:casestudy}
\begin{tabular}{lc|ccccccc}
\toprule
\textbf{Statistics} & \textbf{Full Cohort} & \textbf{Case-cohort} & \textbf{MICE} & \textbf{MICE RSS} & \textbf{MICE ISS} & \textbf{SMC} & \textbf{SMC RSS} & \textbf{SMC ISS} \\
\midrule
Runtime (s) & 1.620 & 0.122 & 38.080 & 1.606 & 2.818 & 27595.268 & 470.314 & 486.819 \\
\specialrule{0.08em}{0.3em}{0.3em}
\multicolumn{2}{c}{\textbf{Log HR}} & \multicolumn{7}{c}{\textbf{Bias from the Full Cohort Estimate}} \\
\addlinespace
$\text{sex}$ & 0.236 & 0.027 & 0.048 & 0.004 & 0.047 & 0.029 & 0.058 & 0.016 \\
$\text{smoke}_1$ & 0.176 & 0.056 & 0.010 & 0.029 & 0.020 & 0.010 & 0.049 & 0.010 \\
$\text{smoke}_2$ & 0.263 & 0.091 & 0.032 & 0.036 & 0.056 & 0.017 & 0.005 & 0.032 \\
$\text{smoke}_3$ & 0.779 & 0.541 & 0.007 & 0.194 & 0.042 & 0.003 & 0.029 & 0.075 \\
$\text{smoke}_4$ & 1.208 & 0.458 & 0.017 & 0.035 & 0.052 & 0.015 & 0.025 & 0.059 \\
$\text{race}_2$ & 0.021 & 0.151 & 0.020 & 0.024 & 0.050 & 0.018 & 0.252 & 0.014 \\
$\text{race}_3$ & -0.010 & 0.684 & 0.003 & 0.116 & 0.040 & 0.015 & 0.106 & 0.034 \\
$\text{race}_4$ & -0.367 & 0.333 & 0.080 & 0.144 & 0.053 & 0.047 & 0.004 & 0.057 \\
$\text{age group}_1$ & 0.698 & 0.076 & 0.000 & 0.028 & 0.004 & 0.005 & 0.014 & 0.006 \\
$\text{age group}_2$ & 1.089 & 0.045 & 0.005 & 0.018 & 0.002 & 0.002 & 0.023 & 0.002 \\
$\text{diabetes}$ & 0.378 & 0.458 & 0.058 & 0.092 & 0.109 & 0.035 & 0.022 & 0.073 \\
$\text{energy}$ & 0.003 & 0.033 & 0.040 & 0.045 & 0.058 & 0.028 & 0.044 & 0.019 \\
$\text{waist}$ & 0.043 & 0.126 & 0.135 & 0.019 & 0.097 & 0.074 & 0.071 & 0.064 \\
$\text{sex}\times\text{waist}$ & 0.066 & 0.053 & 0.058 & 0.062 & 0.013 & 0.014 & 0.001 & 0.006 \\
\specialrule{0.08em}{0.3em}{0.3em}
\multicolumn{2}{c}{\textbf{Standard Error}} & \multicolumn{7}{c}{\textbf{Standard Error}} \\
\addlinespace
$\text{sex}$ & 0.126 & 0.151 & 0.144 & 0.135 & 0.131 & 0.137 & 0.136 & 0.131 \\
$\text{smoke}_1$ & 0.146 & 0.194 & 0.147 & 0.150 & 0.149 & 0.147 & 0.150 & 0.148 \\
$\text{smoke}_2$ & 0.154 & 0.211 & 0.160 & 0.159 & 0.160 & 0.158 & 0.162 & 0.160 \\
$\text{smoke}_3$ & 0.201 & 0.326 & 0.206 & 0.213 & 0.205 & 0.203 & 0.210 & 0.205 \\
$\text{smoke}_4$ & 0.214 & 0.394 & 0.220 & 0.225 & 0.221 & 0.216 & 0.228 & 0.223 \\
$\text{race}_2$ & 0.360 & 0.499 & 0.366 & 0.370 & 0.368 & 0.365 & 0.370 & 0.367 \\
$\text{race}_3$ & 0.452 & 0.643 & 0.456 & 0.458 & 0.456 & 0.453 & 0.459 & 0.455 \\
$\text{race}_4$ & 0.582 & 0.780 & 0.589 & 0.594 & 0.589 & 0.586 & 0.594 & 0.592 \\
$\text{age group}_1$ & 0.168 & 0.187 & 0.169 & 0.170 & 0.171 & 0.169 & 0.170 & 0.170 \\
$\text{age group}_2$ & 0.154 & 0.168 & 0.154 & 0.155 & 0.155 & 0.154 & 0.155 & 0.154 \\
$\text{diabetes}$ & 0.171 & 0.270 & 0.176 & 0.178 & 0.176 & 0.173 & 0.180 & 0.176 \\
$\text{enery}$ & 0.056 & 0.093 & 0.106 & 0.074 & 0.073 & 0.094 & 0.090 & 0.076 \\
$\text{waist}$ & 0.097 & 0.129 & 0.155 & 0.128 & 0.117 & 0.132 & 0.119 & 0.108 \\
$\text{sex}\times\text{waist}$ & 0.115 & 0.156 & 0.158 & 0.158 & 0.129 & 0.146 & 0.121 & 0.130 \\
\bottomrule
\end{tabular}
\caption*{The table presents the estimated log hazard ratios in the \textit{Full Cohort} and their biases relative to the full cohort estimates, along with the corresponding variance estimates. The \textit{Full Cohort} analysis fits a Cox proportional hazards model to the complete dataset and serves as the gold standard. Other methods show absolute bias from the \textit{Full Cohort} estimates. Consistent with the simulation studies, the ISS methods tend to have smaller bias from the full cohort estimates than the RSS methods for both MICE and SMC-FCS. Notably, the standard SMC-FCS requires approximately 7.7 hours, whereas the supersampling methods take about 8 minutes.}
\end{threeparttable}
\end{adjustbox}
\end{table}

\section{Discussion}
\label{sec:discussion}
Building upon the supersampling framework of \citet{borgan2023use}, we proposed an influence function-based supersampling (ISS) approach. Random supersampling (RSS) offers broad applicability when the influence function of the target estimator is unavailable, but it suffers from efficiency loss because it ignores each observation’s contribution to the estimator. When the influence function is obtainable, ISS is designed to reduce the variance of the target estimator by incorporating PPS sampling, balanced sampling, and weight calibration. In our study, the target estimator was the log hazard ratio, whose influence function can be readily computed in R using dfbeta residuals. Consistent with the design features, our simulation results showed that ISS substantially improves efficiency relative to RSS while maintaining a similar computational cost. Notably, even a small supersample size (e.g., 3\% of the total cohort) achieved efficiency comparable to that of imputing the entire cohort outside the case-cohort sample.

The problem of imputing multivariate expensive covariates has become increasingly relevant in modern epidemiological cohorts (\citealp{oluleye2013troponin, deshotels2024vital, fan2018human, meng2025oral}). However, previous methodological work has largely focused on a single expensive covariate. In contrast, we considered scenarios involving multiple expensive biomarkers, where full-cohort imputation is computationally intensive. Our findings suggest that ISS can be incorporated into MI in these settings, indicating its potential applicability to high-dimensional expensive covariates.

Influence function-based sampling has appeared in several recent works (\citealp{ting2018optimal, amorim2021two, shepherd2023multiwave}). \citet{amorim2021two} and \citet{shepherd2023multiwave} focused on constructing validation samples, which are special cases of phase 2 samples, and subsequently performing MI. While \citet{shepherd2023multiwave} adopted an influence function–based sampling strategy, their approach utilized MICE (\citealp{van2011mice}) for imputation. Our work differs in extending MI to the SMC-FCS framework (\citealp{bartlett2015multiple}), which incorporates rejection sampling to eliminate bias arising from interaction or nonlinear terms (\citealp{van2018flexible, keogh2013using}). A key practical distinction is that IF-based phase 2 sampling requires the analysis model to be specified at the design stage. In practice, however, many government health cohorts and research agencies typically select phase 2 samples to support a wide range of future analyses. Our approach complements such practice because it is designed to work with preselected case-cohort samples without requiring the analysis plan to be fixed in advance.

Related approaches by \citet{breslow2009improved, breslow2009using} and \citet{shin2020weight} treated the auxiliary variables as $\bm A_i = \bm{\tilde\psi}_i$, where $\bm{\tilde\psi}_i$ denotes the influence function values computed after a single imputation of $\bm X$. Consistent with these findings, we also observed that adding a calibration equation based on $\bm{\tilde\psi}_i$ further increases efficiency (results not shown). In particular, employing SMC-FCS for single imputation in $\bm X$, consistent with its use in multiple imputation, yielded the most significant efficiency gains.

In \citet{keogh2018multiple}, the authors discussed missing by chance, referring to missing values in low-cost covariates. These cases can also be addressed under our supersampling framework. When the missingness mechanism for $\bm Z$ is missing completely at random (MCAR), it is generally advisable to exclude those units from supersampling. Under MAR, two strategies are possible. One is the exclusion of units with missing values, as in the MCAR case. Although this approach is straightforward, this may induce bias when missingness is substantial. Alternatively, one may impute missing values in $\bm Z$ prior to supersampling using an imputation model that excludes $\bm X$ as predictors. This allows all units to enter the supersampling pool, but the validity of this approach depends heavily on the imputation quality. Therefore, the imputation model should include as many auxiliary variables as possible in the imputation model. Importantly, the imputed values in $\bm Z$ must be reset to missing after supersample selection, because the imputation model used for supersampling deliberately excluded $\bm X$. As noted in previous studies \citep{meng1994multiple, bartlett2015multiple}, excluding analysis variables from the imputation model can introduce bias into the estimator of interest.

\section*{Acknowledgements}
This research was supported by the National Research Foundation of Korea(NRF) Grant funded by the Korea government(MSIT) (RS-2023-00211561) and the Basic Science Research Program through the National Research Foundation of Korea(NRF) funded by the Ministry of Education(RS-2024-00462238).

\bibliographystyle{apalike}
\bibliography{references}

\begin{thebibliography}{}

\bibitem[Amorim et~al., 2021]{amorim2021two}
Amorim, G., Tao, R., Lotspeich, S., Shaw, P.~A., Lumley, T., and Shepherd, B.~E. (2021).
\newblock Two-phase sampling designs for data validation in settings with covariate measurement error and continuous outcome.
\newblock {\em Journal of the Royal Statistical Society: Series A (Statistics in Society)}, 184(4):1368--1389.

\bibitem[Bartlett et~al., 2015]{bartlett2015multiple}
Bartlett, J.~W., Seaman, S.~R., White, I.~R., and Carpenter, J.~R. (2015).
\newblock Multiple imputation of covariates by fully conditional specification: accommodating the substantive model.
\newblock {\em Statistical Methods in Medical Research}, 24(4):462--487.

\bibitem[Beesley et~al., 2016]{beesley2016multiple}
Beesley, L.~J., Bartlett, J.~W., Wolf, G.~T., and Taylor, J.~M. (2016).
\newblock Multiple imputation of missing covariates for the cox proportional hazards cure model.
\newblock {\em Statistics in Medicine}, 35(26):4701--4717.

\bibitem[Borgan et~al., 2023]{borgan2023use}
Borgan, {\O}., Keogh, R.~H., and Nj{\o}s, A. (2023).
\newblock Use of multiple imputation in supersampled nested case-control and case-cohort studies.
\newblock {\em Scandinavian Journal of Statistics}, 50(1):13--37.

\bibitem[Borgan et~al., 2000]{borgan2000exposure}
Borgan, O., Langholz, B., Samuelsen, S.~O., Goldstein, L., and Pogoda, J. (2000).
\newblock Exposure stratified case-cohort designs.
\newblock {\em Lifetime Data Analysis}, 6(1):39--58.

\bibitem[Breslow and Cain, 1988]{breslow1988logistic}
Breslow, N. and Cain, K. (1988).
\newblock Logistic regression for two-stage case-control data.
\newblock {\em Biometrika}, 75(1):11--20.

\bibitem[Breslow et~al., 2009a]{breslow2009improved}
Breslow, N.~E., Lumley, T., Ballantyne, C.~M., Chambless, L.~E., and Kulich, M. (2009a).
\newblock Improved horvitz--thompson estimation of model parameters from two-phase stratified samples: applications in epidemiology.
\newblock {\em Statistics in Biosciences}, 1(1):32--49.

\bibitem[Breslow et~al., 2009b]{breslow2009using}
Breslow, N.~E., Lumley, T., Ballantyne, C.~M., Chambless, L.~E., and Kulich, M. (2009b).
\newblock Using the whole cohort in the analysis of case-cohort data.
\newblock {\em American Journal of Epidemiology}, 169(11):1398--1405.

\bibitem[Chu, 2024]{chu2024}
Chu, C. (2024).
\newblock {\em Joint modeling of longitudinal and survival data: censoring robust estimation, influence function based robust variance and shape based longitudinal clustering}.
\newblock PhD thesis, University of California, Irvine.

\bibitem[Cochran, 1977]{cochran1977sampling}
Cochran, W.~G. (1977).
\newblock {\em Sampling techniques}.
\newblock John Wiley \& Sons, 3rd edition.

\bibitem[Deshotels et~al., 2024]{deshotels2024vital}
Deshotels, M.~R., Al~Rifai, M., Sun, C., Agha, A., Selvin, E., Windham, B.~G., Vaccarino, V., Michos, E.~D., Jneid, H., Levine, G.~N., et~al. (2024).
\newblock Vital exhaustion and biomarkers associated with cardiovascular risk: The aric study.
\newblock {\em JACC: Advances}, 3(11):101355.

\bibitem[Deville and S{\"a}rndal, 1992]{deville1992calibration}
Deville, J.-C. and S{\"a}rndal, C.-E. (1992).
\newblock Calibration estimators in survey sampling.
\newblock {\em Journal of the American Statistical Association}, 87(418):376--382.

\bibitem[Deville and Till{\'e}, 2004]{deville2004efficient}
Deville, J.-C. and Till{\'e}, Y. (2004).
\newblock Efficient balanced sampling: the cube method.
\newblock {\em Biometrika}, 91(4):893--912.

\bibitem[Fan et~al., 2018]{fan2018human}
Fan, X., Alekseyenko, A.~V., Wu, J., Peters, B.~A., Jacobs, E.~J., Gapstur, S.~M., Purdue, M.~P., Abnet, C.~C., Stolzenberg-Solomon, R., Miller, G., et~al. (2018).
\newblock Human oral microbiome and prospective risk for pancreatic cancer: a population-based nested case-control study.
\newblock {\em Gut}, 67(1):120--127.

\bibitem[Folsom et~al., 2002]{folsom2002c}
Folsom, A.~R., Aleksic, N., Catellier, D., Juneja, H.~S., and Wu, K.~K. (2002).
\newblock C-reactive protein and incident coronary heart disease in the atherosclerosis risk in communities (aric) study.
\newblock {\em American Heart Journal}, 144(2):233--238.

\bibitem[Hansen and Hurwitz, 1943]{hansen1943theory}
Hansen, M.~H. and Hurwitz, W.~N. (1943).
\newblock On the theory of sampling from finite populations.
\newblock {\em The Annals of Mathematical Statistics}, 14(4):333--362.

\bibitem[Hughes et~al., 2014]{hughes2014joint}
Hughes, R.~A., White, I.~R., Seaman, S.~R., Carpenter, J.~R., Tilling, K., and Sterne, J.~A. (2014).
\newblock Joint modelling rationale for chained equations.
\newblock {\em BMC Medical Research Methodology}, 14(1):28.

\bibitem[Kalbfleisch and Lawless, 1988]{kalbfleisch1988likelihood}
Kalbfleisch, J. and Lawless, J. (1988).
\newblock Likelihood analysis of multi-state models for disease incidence and mortality.
\newblock {\em Statistics in Medicine}, 7(1-2):149--160.

\bibitem[Keogh et~al., 2018]{keogh2018multiple}
Keogh, R.~H., Seaman, S.~R., Bartlett, J.~W., and Wood, A.~M. (2018).
\newblock Multiple imputation of missing data in nested case-control and case-cohort studies.
\newblock {\em Biometrics}, 74(4):1438--1449.

\bibitem[Keogh and White, 2013]{keogh2013using}
Keogh, R.~H. and White, I.~R. (2013).
\newblock Using full-cohort data in nested case--control and case--cohort studies by multiple imputation.
\newblock {\em Statistics in Medicine}, 32(23):4021--4043.

\bibitem[Kulathinal et~al., 2007]{kulathinal2007case}
Kulathinal, S., Karvanen, J., Saarela, O., and Kuulasmaa, K. (2007).
\newblock Case-cohort design in practice--experiences from the morgam project.
\newblock {\em Epidemiologic Perspectives \& Innovations}, 4:15.

\bibitem[Kulich and Lin, 2004]{kulich2004improving}
Kulich, M. and Lin, D. (2004).
\newblock Improving the efficiency of relative-risk estimation in case-cohort studies.
\newblock {\em Journal of the American Statistical Association}, 99(467):832--844.

\bibitem[Lin and Ying, 1993]{lin1993cox}
Lin, D. and Ying, Z. (1993).
\newblock Cox regression with incomplete covariate measurements.
\newblock {\em Journal of the American Statistical Association}, 88(424):1341--1349.

\bibitem[Liu et~al., 2014]{liu2014stationary}
Liu, J., Gelman, A., Hill, J., Su, Y.-S., and Kropko, J. (2014).
\newblock On the stationary distribution of iterative imputations.
\newblock {\em Biometrika}, 101(1):155--173.

\bibitem[Marti and Chavance, 2011]{marti2011multiple}
Marti, H. and Chavance, M. (2011).
\newblock Multiple imputation analysis of case--cohort studies.
\newblock {\em Statistics in Medicine}, 30(13):1595--1607.

\bibitem[Meng, 1994]{meng1994multiple}
Meng, X.-L. (1994).
\newblock Multiple-imputation inferences with uncongenial sources of input.
\newblock {\em Statistical Science}, 9(4):538--558.

\bibitem[Meng et~al., 2025]{meng2025oral}
Meng, Y., Wu, F., Kwak, S., Wang, C., Usyk, M., Freedman, N.~D., Huang, W.-Y., Um, C.~Y., Gonda, T.~A., Oberstein, P.~E., et~al. (2025).
\newblock Oral bacterial and fungal microbiome and subsequent risk for pancreatic cancer.
\newblock {\em JAMA Oncology}.

\bibitem[Nan, 2004]{nan2004efficient}
Nan, B. (2004).
\newblock Efficient estimation for case-cohort studies.
\newblock {\em Canadian Journal of Statistics}, 32(4):403--419.

\bibitem[Oluleye et~al., 2013]{oluleye2013troponin}
Oluleye, O.~W., Folsom, A.~R., Nambi, V., Lutsey, P.~L., Ballantyne, C.~M., Investigators, A.~S., et~al. (2013).
\newblock Troponin t, b-type natriuretic peptide, c-reactive protein, and cause-specific mortality.
\newblock {\em Annals of Epidemiology}, 23(2):66--73.

\bibitem[Prentice, 1986]{prentice1986case}
Prentice, R.~L. (1986).
\newblock A case-cohort design for epidemiologic cohort studies and disease prevention trials.
\newblock {\em Biometrika}, 73(1):1--11.

\bibitem[Reid and Cr{\'e}peau, 1985]{reid1985influence}
Reid, N. and Cr{\'e}peau, H. (1985).
\newblock Influence functions for proportional hazards regression.
\newblock {\em Biometrika}, 72(1):1--9.

\bibitem[Rubin, 1987]{rubin1987multiple}
Rubin, D.~B. (1987).
\newblock {\em Multiple Imputation for Nonresponse in Surveys}.
\newblock John Wiley \& Sons, New York.

\bibitem[Samuelsen, 1997]{samuelsen1997psudolikelihood}
Samuelsen, S.~O. (1997).
\newblock A pseudolikelihood approach to analysis of nested case-control studies.
\newblock {\em Biometrika}, 84(2):379--394.

\bibitem[Samuelsen et~al., 2007]{samuelsen2007stratified}
Samuelsen, S.~O., {\AA}nestad, H., and Skrondal, A. (2007).
\newblock Stratified case-cohort analysis of general cohort sampling designs.
\newblock {\em Scandinavian Journal of Statistics}, 34(1):103--119.

\bibitem[Schatzkin et~al., 2001]{schatzkin2001design}
Schatzkin, A., Subar, A.~F., Thompson, F.~E., Harlan, L.~C., Tangrea, J., Hollenbeck, A.~R., Hurwitz, P.~E., Coyle, L., Schussler, N., Michaud, D.~S., et~al. (2001).
\newblock Design and serendipity in establishing a large cohort with wide dietary intake distributions: the national institutes of health--american association of retired persons diet and health study.
\newblock {\em American Journal of Epidemiology}, 154(12):1119--1125.

\bibitem[Scheike and Martinussen, 2004]{scheike2004maximum}
Scheike, T.~H. and Martinussen, T. (2004).
\newblock Maximum likelihood estimation for cox's regression model under case-cohort sampling.
\newblock {\em Scandinavian Journal of Statistics}, 31(2):283--293.

\bibitem[Self and Prentice, 1988]{self1988asymptotic}
Self, S.~G. and Prentice, R.~L. (1988).
\newblock Asymptotic distribution theory and efficiency results for case-cohort studies.
\newblock {\em {The Annals of Statistics}}, 16(1):64--81.

\bibitem[Sharp et~al., 2014]{sharp2014review}
Sharp, S.~J., Poulaliou, M., Thompson, S.~G., White, I.~R., and Wood, A.~M. (2014).
\newblock A review of published analyses of case-cohort studies and recommendations for future reporting.
\newblock {\em PlOS ONE}, 9(6):e101176.

\bibitem[Shepherd et~al., 2023]{shepherd2023multiwave}
Shepherd, B.~E., Han, K., Chen, T., Bian, A., Pugh, S., Duda, S.~N., Lumley, T., Heerman, W.~J., and Shaw, P.~A. (2023).
\newblock Multiwave validation sampling for error-prone electronic health records.
\newblock {\em Biometrics}, 79(3):2649--2663.

\bibitem[Shin et~al., 2020]{shin2020weight}
Shin, Y.~E., Pfeiffer, R.~M., Graubard, B.~I., and Gail, M.~H. (2020).
\newblock Weight calibration to improve the efficiency of pure risk estimates from case-control samples nested in a cohort.
\newblock {\em Biometrics}, 76(4):1087--1097.

\bibitem[Stolzenberg-Solomon et~al., 2008]{stolzenberg2008adiposity}
Stolzenberg-Solomon, R.~Z., Adams, K., Leitzmann, M., Schairer, C., Michaud, D.~S., Hollenbeck, A., Schatzkin, A., and Silverman, D.~T. (2008).
\newblock Adiposity, physical activity, and pancreatic cancer in the national institutes of health--aarp diet and health cohort.
\newblock {\em American Journal of Epidemiology}, 167(5):586--597.

\bibitem[Therneau and Li, 1999]{therneau1999computing}
Therneau, T.~M. and Li, H. (1999).
\newblock Computing the cox model for case cohort designs.
\newblock {\em Lifetime Data Analysis}, 5(2):99--112.

\bibitem[Thomas, 1977]{thomas1977appendix}
Thomas, D. (1977).
\newblock Appendix to: Liddell fdk, mc-donald jc, thomas dc. methods of cohort analysis. appraisal by application to asbestos mining (with discussion).
\newblock {\em Journal of the Royal Statistical Society: Series A (General)}, 140(4):469--491.

\bibitem[Till{\'e}, 2011]{tille2011ten}
Till{\'e}, Y. (2011).
\newblock Ten years of balanced sampling with the cube method: an appraisal.
\newblock {\em Survey Methodology}, 37(2):215--226.

\bibitem[Ting and Brochu, 2018]{ting2018optimal}
Ting, D. and Brochu, E. (2018).
\newblock Optimal subsampling with influence functions.
\newblock {\em Advances in Neural Information Processing Systems}, 31:3654--3663.

\bibitem[Tsiatis, 2006]{tsiatis2006semiparametric}
Tsiatis, A.~A. (2006).
\newblock {\em Semiparametric theory and missing data}.
\newblock Springer.

\bibitem[Van~Buuren, 2018]{van2018flexible}
Van~Buuren, S. (2018).
\newblock {\em Flexible imputation of missing data}.
\newblock Chapman and Hall/CRC, 2nd edition.

\bibitem[Van~Buuren and Groothuis-Oudshoorn, 2011]{van2011mice}
Van~Buuren, S. and Groothuis-Oudshoorn, K. (2011).
\newblock mice: Multivariate imputation by chained equations in r.
\newblock {\em Journal of Statistical Software}, 45(3):1--67.

\bibitem[White and Royston, 2009]{white2009imputing}
White, I.~R. and Royston, P. (2009).
\newblock Imputing missing covariate values for the cox model.
\newblock {\em Statistics in Medicine}, 28(15):1982--1998.

\bibitem[White, 1982]{white1982two}
White, J.~E. (1982).
\newblock A two stage design for the study of the relationship between a rare exposure and a rare disease.
\newblock {\em American Journal of Epidemiology}, 115(1):119--128.

\bibitem[Xie and Meng, 2017]{xie2017dissecting}
Xie, X. and Meng, X. (2017).
\newblock Dissecting multiple imputation from a multi-phase inference perspective: What happens when god's, imputer's and analyst's models are uncongenial?
\newblock {\em {Statistica Sinica}}, 27(4):1485--1594.

\end{thebibliography}

\appendix
\section*{Appendix}
\section{Stratified Case-Cohort Simulation Studies}
The simulation results for the stratified case-cohort studies are presented. 
\begin{table}[htbp]
\centering
\begin{adjustbox}{max width=\textwidth}
\begin{threeparttable}
\captionsetup{font=large, labelfont=bf, labelsep=period}
\caption{Simulation Results: Analysis Model Without an Interaction Term}
\begin{tabular}{lcccccccc}
\toprule
\textbf{Statistics} & \textbf{Full Cohort} & \textbf{Case-cohort} & \textbf{MICE} & \textbf{MICE RSS} & \textbf{MICE ISS} & \textbf{SMC} & \textbf{SMC RSS} & \textbf{SMC ISS} \\
\midrule
avg.time & 0.398 & 0.023 & 38.476 & 4.241 & 4.269 & 1763.246 & 71.974 & 83.022 \\
max time & 1.015 & 0.065 & 48.565 & 5.597 & 5.523 & 1884.327 & 86.289 & 107.282 \\
min time & 0.271 & 0.014 & 32.106 & 3.486 & 3.541 & 1579.234 & 41.814 & 38.755 \\
\specialrule{0.08em}{0.3em}{0.3em}
bias $\mathrm{z}_1$ & 0.001 & 0.251 & 0.031 & 0.106 & 0.024 & 0.053 & 0.120 & 0.048 \\
bias $\mathrm{z}_2$ & 0.010 & 0.085 & 0.010 & 0.034 & 0.014 & 0.020 & 0.039 & 0.027 \\
bias $\mathrm{z}_{3.2}$ & 0.001 & 0.026 & 0.003 & 0.009 & 0.001 & 0.004 & 0.019 & 0.002 \\
bias $\mathrm{z}_{3.3}$ & 0.005 & 0.027 & 0.003 & 0.001 & 0.006 & 0.001 & 0.008 & 0.002 \\
bias $\mathrm{xc}_1$ & 0.001 & 0.068 & 0.023 & 0.002 & 0.017 & 0.003 & 0.019 & 0.004 \\
bias $\mathrm{xc}_2$ & 0.002 & 0.023 & 0.006 & 0.001 & 0.002 & 0.001 & 0.004 & 0.002 \\
bias $\mathrm{xc}_3$ & 0.001 & 0.013 & 0.004 & 0.003 & 0.001 & 0.001 & 0.005 & 0.003 \\
bias $\mathrm{xc}_4$ & 0.001 & 0.015 & 0.006 & 0.000 & 0.005 & 0.001 & 0.003 & 0.002 \\
bias $\mathrm{xb}_1$ & 0.001 & 0.027 & 0.005 & 0.011 & 0.017 & 0.024 & 0.019 & 0.027 \\
bias $\mathrm{xb}_2$ & 0.000 & 0.061 & 0.014 & 0.029 & 0.032 & 0.039 & 0.046 & 0.040 \\
\specialrule{0.08em}{0.3em}{0.3em}
mc.se $\mathrm{z}_1$ & 0.067 & 0.207(10.4\%) & 0.075(79.8\%) & 0.134(24.8\%) & 0.076(76.9\%) & 0.079(70.9\%) & 0.137(23.6\%) & 0.081(66.9\%) \\
mc.se $\mathrm{z}_2$ & 0.128 & 0.380(11.3\%) & 0.156(67.6\%) & 0.223(32.8\%) & 0.157(66.3\%) & 0.159(64.6\%) & 0.219(34.1\%) & 0.163(61.6\%) \\
mc.se $\mathrm{z}_{3.2}$ & 0.146 & 0.421(12.1\%) & 0.179(66.7\%) & 0.250(34.3\%) & 0.180(66.0\%) & 0.182(64.3\%) & 0.255(32.9\%) & 0.188(60.5\%) \\
mc.se $\mathrm{z}_{3.3}$ & 0.154 & 0.446(12.0\%) & 0.187(68.0\%) & 0.271(32.3\%) & 0.188(67.7\%) & 0.190(65.7\%) & 0.261(34.9\%) & 0.194(63.5\%) \\
mc.se $\mathrm{xc}_1$ & 0.063 & 0.192(10.9\%) & 0.130(23.6\%) & 0.137(21.3\%) & 0.130(23.6\%) & 0.137(21.4\%) & 0.146(18.9\%) & 0.142(20.0\%) \\
mc.se $\mathrm{xc}_2$ & 0.064 & 0.197(10.7\%) & 0.132(23.9\%) & 0.144(19.9\%) & 0.134(23.0\%) & 0.140(21.2\%) & 0.149(18.6\%) & 0.141(20.9\%) \\
mc.se $\mathrm{xc}_3$ & 0.062 & 0.188(10.8\%) & 0.129(23.0\%) & 0.138(20.3\%) & 0.132(22.1\%) & 0.135(21.1\%) & 0.145(18.3\%) & 0.138(20.1\%) \\
mc.se $\mathrm{xc}_4$ & 0.060 & 0.200(9.0\%) & 0.131(21.2\%) & 0.144(17.4\%) & 0.134(20.1\%) & 0.139(18.7\%) & 0.149(16.3\%) & 0.143(17.8\%) \\
mc.se $\mathrm{xb}_1$ & 0.128 & 0.378(11.5\%) & 0.265(23.3\%) & 0.284(20.3\%) & 0.260(24.3\%) & 0.291(19.4\%) & 0.302(18.0\%) & 0.292(19.3\%) \\
mc.se $\mathrm{xb}_2$ & 0.140 & 0.394(12.6\%) & 0.282(24.5\%) & 0.297(22.2\%) & 0.271(26.6\%) & 0.294(22.5\%) & 0.311(20.2\%) & 0.302(21.4\%) \\
\specialrule{0.08em}{0.3em}{0.3em}
est.se $\mathrm{z}_1$ & 0.067(0.942) & 0.138(0.536) & 0.082(0.954) & 0.129(0.857) & 0.094(0.981) & 0.083(0.915) & 0.128(0.824) & 0.094(0.954) \\
est.se $\mathrm{z}_2$ & 0.130(0.956) & 0.278(0.845) & 0.166(0.973) & 0.233(0.960) & 0.180(0.983) & 0.167(0.961) & 0.232(0.963) & 0.181(0.972) \\
est.se $\mathrm{z}_{3.2}$ & 0.150(0.957) & 0.327(0.872) & 0.190(0.967) & 0.270(0.957) & 0.205(0.973) & 0.190(0.964) & 0.269(0.969) & 0.208(0.972) \\
est.se $\mathrm{z}_{3.3}$ & 0.157(0.947) & 0.339(0.867) & 0.197(0.959) & 0.279(0.959) & 0.213(0.977) & 0.197(0.956) & 0.279(0.963) & 0.215(0.974) \\
est.se $\mathrm{xc}_1$ & 0.063(0.951) & 0.145(0.839) & 0.131(0.946) & 0.159(0.974) & 0.143(0.960) & 0.130(0.937) & 0.155(0.962) & 0.141(0.946) \\
est.se $\mathrm{xc}_2$ & 0.063(0.945) & 0.147(0.847) & 0.133(0.952) & 0.158(0.973) & 0.142(0.957) & 0.132(0.930) & 0.156(0.964) & 0.142(0.948) \\
est.se $\mathrm{xc}_3$ & 0.063(0.951) & 0.146(0.869) & 0.132(0.950) & 0.157(0.968) & 0.142(0.963) & 0.131(0.945) & 0.156(0.963) & 0.140(0.954) \\
est.se $\mathrm{xc}_4$ & 0.063(0.961) & 0.145(0.839) & 0.131(0.947) & 0.154(0.965) & 0.139(0.961) & 0.128(0.935) & 0.153(0.953) & 0.139(0.942) \\
est.se $\mathrm{xb}_1$ & 0.129(0.955) & 0.297(0.871) & 0.268(0.954) & 0.316(0.967) & 0.278(0.961) & 0.265(0.922) & 0.315(0.959) & 0.283(0.947) \\
est.se $\mathrm{xb}_2$ & 0.135(0.945) & 0.304(0.862) & 0.270(0.949) & 0.316(0.965) & 0.280(0.952) & 0.268(0.925) & 0.317(0.945) & 0.286(0.940) \\
\bottomrule
\end{tabular}
\end{threeparttable}
\end{adjustbox}
\end{table}

\begin{table}[htbp]
\centering
\begin{adjustbox}{max width=\textwidth}
\begin{threeparttable}
\captionsetup{font=large, labelfont=bf, labelsep=period}
\caption{Simulation Results: Analysis Model With an Interaction Term}
\begin{tabular}{lcccccccc}
\toprule
\textbf{Statistics} & \textbf{Full Cohort} & \textbf{Case-cohort} & \textbf{MICE} & \textbf{MICE RSS} & \textbf{MICE ISS} & \textbf{SMC} & \textbf{SMC RSS} & \textbf{SMC ISS} \\
\midrule
avg.time & 0.405 & 0.025 & 39.697 & 4.330 & 4.378 & 1955.004 & 76.493 & 97.681 \\
max time & 0.921 & 0.409 & 48.540 & 5.105 & 5.207 & 2077.359 & 86.546 & 123.997 \\
min time & 0.320 & 0.015 & 33.854 & 3.617 & 3.815 & 1741.800 & 44.382 & 44.820 \\
\specialrule{0.08em}{0.3em}{0.3em}
bias $\mathrm{z}_1$ & 0.006 & 0.240 & 0.086 & 0.136 & 0.077 & 0.070 & 0.119 & 0.063 \\
bias $\mathrm{z}_2$ & 0.009 & 0.064 & 0.030 & 0.013 & 0.018 & 0.007 & 0.023 & 0.015 \\
bias $\mathrm{z}_{3.2}$ & 0.000 & 0.008 & 0.007 & 0.000 & 0.012 & 0.000 & 0.006 & 0.002 \\
bias $\mathrm{z}_{3.3}$ & 0.001 & 0.060 & 0.021 & 0.001 & 0.022 & 0.002 & 0.019 & 0.003 \\
bias $\mathrm{xc}_1$ & 0.000 & 0.152 & 0.165 & 0.104 & 0.189 & 0.054 & 0.061 & 0.077 \\
bias $\mathrm{xc}_2$ & 0.000 & 0.014 & 0.013 & 0.003 & 0.008 & 0.004 & 0.004 & 0.003 \\
bias $\mathrm{xc}_3$ & 0.002 & 0.005 & 0.012 & 0.005 & 0.009 & 0.006 & 0.001 & 0.003 \\
bias $\mathrm{xc}_4$ & 0.001 & 0.011 & 0.008 & 0.000 & 0.008 & 0.000 & 0.002 & 0.003 \\
bias $\mathrm{xb}_1$ & 0.003 & 0.016 & 0.011 & 0.002 & 0.003 & 0.012 & 0.013 & 0.012 \\
bias $\mathrm{xb}_2$ & 0.007 & 0.040 & 0.008 & 0.021 & 0.022 & 0.043 & 0.040 & 0.042 \\
bias $\mathrm{z}_1\mathrm{xc}_1$ & 0.004 & 0.024 & 0.186 & 0.084 & 0.188 & 0.049 & 0.016 & 0.051 \\
\specialrule{0.08em}{0.3em}{0.3em}
mc.se $\mathrm{z}_1$ & 0.088 & 0.265(10.9\%) & 0.097(81.2\%) & 0.160(30.1\%) & 0.100(76.6\%) & 0.099(77.8\%) & 0.154(32.2\%) & 0.102(74.3\%) \\
mc.se $\mathrm{z}_2$ & 0.125 & 0.405(9.5\%) & 0.167(56.3\%) & 0.242(26.6\%) & 0.172(52.5\%) & 0.178(49.1\%) & 0.244(26.3\%) & 0.183(46.8\%) \\
mc.se $\mathrm{z}_{3.2}$ & 0.149 & 0.463(10.3\%) & 0.193(59.1\%) & 0.292(25.9\%) & 0.201(54.5\%) & 0.203(53.6\%) & 0.285(27.2\%) & 0.206(51.8\%) \\
mc.se $\mathrm{z}_{3.3}$ & 0.155 & 0.492(9.9\%) & 0.202(58.9\%) & 0.295(27.6\%) & 0.210(54.7\%) & 0.214(52.5\%) & 0.298(27.1\%) & 0.221(49.1\%) \\
mc.se $\mathrm{xc}_1$ & 0.117 & 0.272(18.7\%) & 0.143(67.7\%) & 0.174(45.7\%) & 0.156(57.1\%) & 0.154(58.1\%) & 0.180(42.4\%) & 0.169(48.4\%) \\
mc.se $\mathrm{xc}_2$ & 0.065 & 0.201(10.6\%) & 0.136(23.0\%) & 0.148(19.5\%) & 0.139(22.1\%) & 0.146(20.1\%) & 0.150(19.0\%) & 0.148(19.4\%) \\
mc.se $\mathrm{xc}_3$ & 0.064 & 0.200(10.4\%) & 0.136(22.5\%) & 0.148(18.9\%) & 0.136(22.4\%) & 0.148(19.0\%) & 0.150(18.6\%) & 0.148(19.0\%) \\
mc.se $\mathrm{xc}_4$ & 0.065 & 0.202(10.4\%) & 0.133(23.9\%) & 0.143(20.8\%) & 0.134(23.6\%) & 0.139(22.1\%) & 0.149(19.2\%) & 0.141(21.3\%) \\
mc.se $\mathrm{xb}_1$ & 0.129 & 0.408(9.9\%) & 0.283(20.7\%) & 0.305(17.8\%) & 0.268(23.0\%) & 0.300(18.4\%) & 0.311(17.1\%) & 0.308(17.5\%) \\
mc.se $\mathrm{xb}_2$ & 0.134 & 0.425(10.0\%) & 0.281(22.7\%) & 0.308(19.0\%) & 0.275(23.9\%) & 0.301(19.8\%) & 0.316(18.0\%) & 0.305(19.4\%) \\
mc.se $\mathrm{z}_1\mathrm{xc}_1$ & 0.069 & 0.171(16.1\%) & 0.074(86.4\%) & 0.099(48.1\%) & 0.077(80.0\%) & 0.085(64.4\%) & 0.102(45.0\%) & 0.090(58.3\%) \\
\specialrule{0.08em}{0.3em}{0.3em}
est.se $\mathrm{z}_1$ & 0.086(0.942) & 0.170(0.618) & 0.106(0.901) & 0.165(0.870) & 0.118(0.955) & 0.100(0.903) & 0.155(0.887) & 0.114(0.952) \\
est.se $\mathrm{z}_2$ & 0.129(0.948) & 0.276(0.823) & 0.184(0.966) & 0.277(0.978) & 0.207(0.982) & 0.178(0.951) & 0.268(0.968) & 0.206(0.977) \\
est.se $\mathrm{z}_{3.2}$ & 0.148(0.956) & 0.324(0.838) & 0.209(0.963) & 0.322(0.968) & 0.238(0.985) & 0.203(0.958) & 0.312(0.963) & 0.235(0.978) \\
est.se $\mathrm{z}_{3.3}$ & 0.158(0.956) & 0.339(0.830) & 0.220(0.968) & 0.338(0.972) & 0.247(0.975) & 0.215(0.947) & 0.325(0.967) & 0.245(0.967) \\
est.se $\mathrm{xc}_1$ & 0.122(0.956) & 0.214(0.849) & 0.167(0.900) & 0.227(0.977) & 0.216(0.971) & 0.166(0.964) & 0.213(0.975) & 0.215(0.985) \\
est.se $\mathrm{xc}_2$ & 0.063(0.940) & 0.141(0.823) & 0.140(0.955) & 0.173(0.974) & 0.151(0.961) & 0.131(0.933) & 0.164(0.961) & 0.145(0.948) \\
est.se $\mathrm{xc}_3$ & 0.063(0.943) & 0.142(0.832) & 0.139(0.956) & 0.174(0.977) & 0.152(0.968) & 0.133(0.913) & 0.165(0.965) & 0.147(0.961) \\
est.se $\mathrm{xc}_4$ & 0.063(0.939) & 0.140(0.829) & 0.137(0.953) & 0.172(0.975) & 0.149(0.961) & 0.130(0.931) & 0.162(0.966) & 0.143(0.954) \\
est.se $\mathrm{xb}_1$ & 0.129(0.955) & 0.292(0.833) & 0.282(0.960) & 0.353(0.980) & 0.298(0.969) & 0.266(0.928) & 0.334(0.970) & 0.295(0.947) \\
est.se $\mathrm{xb}_2$ & 0.135(0.953) & 0.300(0.843) & 0.284(0.955) & 0.357(0.980) & 0.300(0.964) & 0.268(0.909) & 0.339(0.972) & 0.297(0.946) \\
est.se $\mathrm{z}_1\mathrm{xc}_1$ & 0.069(0.946) & 0.118(0.832) & 0.099(0.581) & 0.160(0.985) & 0.118(0.745) & 0.084(0.915) & 0.131(0.983) & 0.107(0.968) \\
\bottomrule
\end{tabular}
\end{threeparttable}
\end{adjustbox}
\end{table}

\section{Case Study}
The table summarizing the strata and the number of participants in the NIH-AARP Diet and Health Study is presented.

\begin{table}[htbp]
\centering
\captionsetup{font=small, labelfont=bf, labelsep=period}
\caption{Stratified Sampling Design for the NIH--AARP Diet and Health Study}
\label{tab:strata_summary}
\resizebox{\textwidth}{!}{
\begin{tabular}{lcccccc}
\toprule
Strata & Male 50-59 & Male 60-64 & Male 65-71 & Female 50-59 & Female 60-64 & Female 65-71 \\
\midrule
Full cohort ($N_h$) & 45,807 & 41,459 & 56,725 & 31,049 & 25,418 & 31,995 \\
Cases ($D_h$) & 40 & 66 & 125 & 16 & 30 & 60 \\
Cases in subcohort ($d_h$) & 0 & 1 & 1 & 0 & 0 & 0 \\
Subcohort ($n_{sc,h}$) & 200 & 200 & 250 & 130 & 90 & 130 \\
Supersample ($n_{1h}$) & 400 & 400 & 500 & 250 & 200 & 250 \\
\bottomrule
\end{tabular}
}
\end{table}

\end{document}